\documentclass[12pt]{article}
\usepackage{amsmath} 
\usepackage{amsfonts}
\usepackage{amssymb,graphics,psfrag}
\usepackage[pdftex]{color,graphicx}
\usepackage{array,epsfig}
\usepackage{hyperref}

\usepackage{pgf}
\usepackage{tikz}
\usepackage{pgf}
\usepackage{pgffor}
\usepgfmodule{shapes}
\usepgfmodule{plot}
\usetikzlibrary{decorations}
\usetikzlibrary{arrows}
\usetikzlibrary{snakes}
\usetikzlibrary{shapes}

\tikzstyle{decision} = [diamond, draw, 
    text width=4.5em, text badly centered, node distance=3cm, inner sep=0pt]
\tikzstyle{block} = [rectangle, draw, 
    text width=5em, text centered, rounded corners, minimum height=4em]
\tikzstyle{line} = [draw, -latex']
\tikzstyle{cloud} = [draw, ellipse,fill=red!20, node distance=3cm,
    minimum height=2em]

\def\hybrid{\topmargin -20pt    \oddsidemargin 0pt
        \headheight 0pt \headsep 0pt
        \textwidth 6.25in       
        \textheight 9 in       
        \marginparwidth .875in
        \parskip 5pt plus 1pt 
          \jot = 1.5ex
   }
\hybrid
\numberwithin{equation}{section}
\numberwithin{table}{section}

\newcommand{\beq}{\begin{equation}}
\newcommand{\eeq}{\end{equation}}
\newcommand{\be}{\begin{equation}}
\newcommand{\ee}{\end{equation}}
\newcommand{\bea}{\begin{eqnarray}}
\newcommand{\eea}{\end{eqnarray}}   
\newcommand{\ben}{\begin{eqnarray*}}
\newcommand{\een}{\end{eqnarray*}}                  
\newcommand{\ba}{\begin{aligned}}
\newcommand{\ea}{\end{aligned}}
\newcommand{\bt}{\begin{tabular}}
\newcommand{\et}{\end{tabular}}
\newcommand{\bc}{\begin{center}}
\newcommand{\ec}{\end{center}}

\newcommand{\cC}{\mathcal{C}}
\newcommand{\cD}{\mathcal{D}}

\newcommand{\cA}{\mathcal{A}}

\newcommand{\cB}{\mathcal{B}}

\newcommand{\cref}{{\bf [check ref]}}

\def\a{{$\alpha'$\,}}

\psfrag{n1}{$\nu_1$}
\psfrag{n2}{$\nu_2$}
\psfrag{n1'}{$\nu_1'$}
\psfrag{n2'}{$\nu_2'$}
\psfrag{n9}{$\nu_9$}
\psfrag{n10'}{$\nu_{10}'$}
\psfrag{t1}{$\tilde{\nu}_1$}
\psfrag{t2}{$\tilde{\nu}_2$}
\psfrag{t9}{$\tilde{\nu}_9$}
\psfrag{t1'}{$\tilde{\nu}_1'$}
\psfrag{t2'}{$\tilde{\nu}_2'$}
\psfrag{t10'}{$\tilde{\nu}_{10}'$}

\def\blfootnote{\xdef\@thefnmark{}\@footnotetext} 
\long\def\symbolfootnote[#1]#2{\begingroup%
\def\thefootnote{\fnsymbol{footnote}}\footnote[#1]{#2}\endgroup}

\begin{document}

\baselineskip=17pt

\begin{titlepage}
\titlepage
\begin{flushright}{CERN-PH-TH/2012-063 \\
 MPP-2012-62}
\end{flushright}
\vskip 2.5cm
{ \bf \huge On Flux Quantization in F-Theory II: \\  \\ Unitary and Symplectic Gauge Groups}
\vskip 2.3cm

\begin{center}
{\bf \large Andr\'es Collinucci}
\vskip 0.3cm
\em 
Theory Group, Physics Department, CERN CH-1211 Geneva 23, Switzerland\,,\\ \&\\

Physique Th\'eorique et Math\'ematique Universit\'e Libre de Bruxelles, C.P. 231, 1050
Bruxelles, Belgium

\vskip 0.5 cm

{\bf \large Raffaele Savelli}
\vskip 0.3cm

Max-Planck-Institut f\"ur Physik, \\
F\"ohringer Ring 6, 80805 Munich, Germany

\vskip 0.7cm

\large \bf Abstract
\end{center}

\normalsize 
We study the quantization of the M-theory G-flux on elliptically fibered Calabi-Yau fourfolds with singularities giving rise to unitary and symplectic gauge groups. We seek and find its relation to the Freed-Witten quantization of worldvolume fluxes on 7-branes in type IIB orientifold compactifications on Calabi-Yau threefolds. By explicitly constructing the appropriate four-cycles on which to calculate the periods of the second Chern class of the fourfolds, we find that there is a half-integral shift in the quantization of $G$-flux whenever the corresponding dual 7-brane is wrapped on a non-spin submanifold.
This correspondence of quantizations holds for all unitary and symplectic gauge groups, except for $SU(3)$, which behaves mysteriously.\\
We also perform our analysis in the case where, in addition to the aforementioned gauge groups, there is also a `flavor' $U(1)$-gauge group.

\vskip2cm

\vskip1.5\baselineskip

\vfill
\vskip 2.mm
\end{titlepage}

\tableofcontents

\newpage

%
%

\section{Introduction}

The investigation of global aspects of F-theory compactifications has received much attention in the past few years \cite{Collinucci:2008zs, Collinucci:2009uh, Blumenhagen:2009up, Blumenhagen:2009yv, Grimm:2009yu, Marsano:2009ym, Chen:2010ts, Esole:2011sm, Esole:2011cn}. Although primarily driven by phenomenological aspirations, these studies have furthered our understanding of fundamental aspects in general F-theory configurations. One such notable aspect concerns G-flux \cite{Marsano:2011nn, Marsano:2010ix, Grimm:2009ef, Grimm:2011tb, Braun:2011zm}. A detailed knowledge of these objects and of their properties is crucial for both model building issues, like generation of chiral matter in the gauge theory sector and moduli stabilization in the gravitational sector, and more fundamental questions regarding the quantum consistency of the UV theory. An essential tool which has driven the recent development of the field is the duality between F and M-theory \cite{Vafa:1996xn}, which allows for a rigorous and elegant definition of F-theory through its long-wavelength limit (eleven-dimensional supergravity) (for reviews see \cite{Denef:2008wq, Weigand:2010wm}).

However, many questions remain unanswered. In this paper we will address the the quantization rule for the M-theory $G_4$ flux \cite{Witten:1996md} in its relation to the Freed-Witten quantization condition for gauge fluxes on 7-brane configurations \cite{Minasian:1997mm, Freed:1999vc}. In \cite{Collinucci:2010gz}, the authors studied the link between the half-integral shifts in the quantization condition of $G_4$ and in the quantization condition of the dual 7-brane gauge flux. While proving that F-theory configurations with smooth Weierstrass representation never display such a shift, they were able to indirectly analyze the singular cases. Indeed, they matched the quantization conditions of $G_4$ and of its dual 7-brane flux by using the type IIB weak coupling limit of F-theory, and comparing the M2 and D3 induced charges. However, the analysis focused on symplectic gauge groups, which are more tractable at weak coupling. 

In this paper, we generalize our previous studies, and render them physically more intuitive. We will in fact find a purely M-theoretic way of detecting the membrane anomaly. Moreover we will show that our method provides a direct lift to M-theory of the usual strategy adopted to detect Freed-Witten anomaly in perturbative D-brane physics. More precisely, a D7-brane wrapping a non-spin manifold carries a half-integrally quantized flux. Given the 2-cycle on which the period of the flux is half-integral, we identify the corresponding 4-cycle in M-theory where the period of the second Chern class of the elliptic fourfold is an odd integer, thus modifying the quantization rule of $G_4$. 

The authors of \cite{Krause:2012yh} have shown that no shift in the quantization of $G_4$ can be detected by integrating $c_2$ on holomorphic 4-cycles given by complete intersections of divisors, or on fibrations of exceptional $\mathbb{P}^1$'s over matter curves. Our strategy will be to look at holomorphic 4-cycles that are not complete intersections of divisors with the Calabi-Yau hypersurface, much in the same spirit of \cite{Braun:2011zm}. Their explicit description in terms of sets of algebraic equations is not present in a general point of the moduli space of the fourfold.

By mildly constraining the complex structure of the Weierstrass model, we are able to find those 4-cycles, which are generically not holomorphic, and therefore geometrically not tractable. In fact, we will propose two alternative methods to perform such a constraint, based on two different Ans\"atze on the coefficients of the `Tate model', both of them with a clear physical interpretation in type IIA string theory. One of them allows us to make a more direct treatment of the symplectic series, hence extending and clarifying the results of \cite{Collinucci:2010gz}, while the other one gives us the possibility to extend the same analysis to the whole tower of unitary gauge groups. This is indeed a very important generalization in light of the promising Grand-Unified-Theory models which are mainly based on the existence of an $SU(5)$ gauge stack of 7-branes. 

A notable exception in our investigation is represented by the $SU(3)$ configuration, for which neither of our two prescriptions work. We will present the details of this model in an appendix and explain the reasons of this anomalous behavior. Finally we will extend our analysis to models with flavor $U(1)$-gauge groups, whose relevance in the generation of chiral matter on D-brane intersections is well-known \cite{Grimm:2010ez}.

The paper is organized as follows: After summarizing our results in section \ref{sec:summary}, we address the $Sp$-series, finding appropriate four-cycles on which to detect the half-integral shift in the quantization of $G_4$. We relate these directly to the two-cycles on which the Freed-Witten shift of the corresponding D7 gauge flux is detected.\\
In section \ref{sec:suseries}, we proceed with the case of unitary groups, which are generically less tractable from the type IIB point of view. Nevertheless, we find that the quantization of $G_4$ still mimics that of gauge flux of the corresponding D7-stack. To make our formulae more palatable, we work out explicitly the interesting case of $SU(5)$.\\
In section \ref{sec:u1restrictions} we take our investigation one step further by adding singularities that give rise to `flavor' $U(1)$ gauge groups. We conclude with an outlook in section \ref{outlook}.\\
In appendix \ref{GeneralSU} we give a general description of the resolution of CY fourfolds with arbitrary $SU(N)$ singularities over arbitrary base threefolds.\\
In appendix \ref{SU(3)} we say a few words about the special case of $SU(3)$ gauge group.

\section{Summary of results} \label{sec:summary}

Let us collect here the main results of this paper for the convenience of the reader.

We have found non-complete intersection, integral 4-cycles of the F-theory fourfold on which we are able to measure odd values of the second Chern class and hence half-integral shifts in the quantization condition of $G_4$. We have performed this analysis for both the symplectic and the unitary series of gauge groups. We find half-integral shifts precisely when the corresponding 7-brane stack is wrapping a non-spin manifold, thus directly connecting this topological effect with the Freed-Witten quantization condition in type II string theory.

For a 7-brane stack with $Sp(N)$ gauge group wrapped on a divisor $\mathcal{D}$ that contains a holomorphic curve $\mathcal{C}$, we explicitly construct a holomorphic surface $C_{(4)}$ in the resolved F-theory CY fourfold $\tilde Z_4$, such that

\bea
\int_{C_{(4)}}c_2(\tilde Z_4) &=&\int_{\mathcal{C}}\left[7\,c_1(B_3)-(2N-1)\,\mathcal{D}\right]\,.
\eea

This 4-cycle has the geometry of a $\mathbb{P}^1$ fibered over the curve $\mathcal{C} \subset \mathcal{D}$ and it lifts a loop of type IIA open strings stretching between one D-brane of the stack and the Whitney umbrella D-brane. Hence, if $\mathcal{D}$ is not spin, such that its first Chern class has an odd period on $\mathcal{C}$, then $c_2(\tilde Z_4)$ will have an odd period on our 4-cycle $C_{(4)}$.

For the $SU(2N)$-series with $N\geq2$ we construct the 4-cycle in the same way. The general formula for the integrated second Chern class is now
\bea 
\int_{C_{(4)}} c_2(\tilde Z_4) &=&\int_{\mathcal{C}} \left( 6\,c_1(B_3)-(2\,N-1) \cdot \mathcal{D} \right)\,.
\eea

For the $SU(2N+1)$-series with $N\geq2$ we construct a different type of 4-cycle $C_{(4)}'$, however it also consists in fibering a $\mathbb{P}^1$ over $\mathcal{C}$. 
Physically, this other 4-cycle is lifting a loop of closed, non-orientable type IIA open strings. The general result for the integrated second Chern class is
\bea
\int_{C_{(4)}^{\prime}}c_2(\tilde{Z}_4)&=&\int_{\mathcal{C}} \mathcal{D}\,.
\eea
It is worth remarking that this second strategy works fine also for $SU(2N)$ with $N\geq3$.

We finally have investigated flux quantization in F-theory models with the so-called $U(1)$-restriction \cite{Grimm:2010ez}. For gauge groups $SU(N\geq 5)$ nothing changes with respect to the previous discussion, if we treat all of these cases with the second type of 4-cycle, $C_{(4)}'$. 

For the $Sp$-series we conjecture that odd-rank $Sp$ groups lead to even second Chern classes and viceversa. This is because in the odd-rank cases the two branches of the `flavour brane' (shaped like a \emph{Whitney umbrella}, as shown in \cite{Collinucci:2008pf}) are separately non-spin and the induced flux on them is also half-integral. This effect cancels the analog effect arising on the non-abelian stack. In contrast,  in the even-rank cases, the branches of the `flavour brane' are spin and the induced gauge flux is integrally quantized. For this reason we still find, by explicit computation, an odd second Chern class also for the $U(1)$-restricted $SU(4)$-model.

\section{Detection of the Freed-Witten anomaly in F-theory: Sp-series}

In \cite{Collinucci:2010gz} it has been argued that, for M-theory on elliptically fibered Calabi-Yau fourfolds with a codimension one singularity of  Kodaira type I$_{2N}^{ns}$, the $G_4$ flux is quantized in terms of half-integers if and only if the singular locus does not admit spin structures. This is the F-theory counterpart of the Freed-Witten quantization condition for the gauge flux on the D7-brane $Sp(N)$-stack\footnote{We will always use for symplectic groups the notation where $N$ refers to the rank.} which wraps a non-spin submanifold of the Calabi-Yau threefold in the weakly coupled type IIB string picture. However, it has not been clarified yet on which 4-cycles of the fourfold one is actually able to detect the shifted quantization condition. In fact, it turns out that such cycles are not manifest as algebraic submanifolds in a generic point of the moduli space of the fourfold. In the same spirit as in \cite{Braun:2011zm}, in this section we will identify them by mildly constraining the complex structure moduli of the fourfold. As we will see, the way to perform the restriction is suggested by the description of the system in the weak coupling limit, which for this reason we will present first.

Throughout this paper, we will use the phrase \emph{Freed-Witten anomaly}, to refer to the half-integral shift in the quantization of worldvolume flux, as opposed to the anomaly related to $H_3$-fluxes that restrict non-trivially to the brane.

\subsection{Sen limit}\label{SenLimitFW}

As reviewed in \cite{Collinucci:2010gz}, Sen's weak coupling limit of an F-theory configuration is described in terms of a Calabi-Yau threefold equipped with an orientifold involution. If $B_3$ is the base manifold of the F-theory elliptic fourfold, the type IIB Calabi-Yau threefold is the double cover of $B_3$ defined by the following equation 
\bea\label{genericCY3}
X_3&:&(\xi-a_1)(\xi+a_1)=4a_2\,.
\eea
in an ambient four-dimensional manifold. The ambient variety is obtained by adding to $B_3$ the homogeneous coordinate $\xi$ of degree $c_1(B_3)$. The polynomials $a_{1,2}$, which depend on the coordinates of $B_3$, are sections of $K_{B_3}^{-1}$, $K_{B_3}^{-2}$ respectively (they are the first Tate coefficients of the elliptic fourfold defining equation, see eq. \eqref{singularZ4} below). The geometrical action of the orientifold involution is defined by the operation $\xi\to -\xi$ and the equation $\xi=0$ determines the locus wrapped by an O7$^-$-plane. 

If the F-theory fourfold has an $Sp(N)$ singularity over the locus $\{D=0\}\subset B_3$ with $D$ an homogeneous polynomial, in the weak coupling limit there is a stack of $2N$  orientifold-invariant D7-branes wrapping the manifold given by the complete intersection
\bea\label{invariantS2}
S_2&:&\left\{\begin{array}{l}D=0\\ (\xi-a_1)(\xi+a_1)=4a_2\;.\end{array}\right.
\eea
Being transverse to the O7-plane, the stack accommodates an $Sp(N)$ gauge theory. Consider the case in which the submanifold \eqref{invariantS2} is non-spin. Then, there must exist 2-cycles of the D7-stack on which $c_1(S_2)$ has an odd period. However, these may have no algebraic representatives in a generic region of the moduli space of the D7-divisor. That is indeed the case here, and the mildest assumption we can make in order that those Riemann surfaces show up is as follows. Let us constrain the polynomials $D$ and $a_2$ to be of the form
\bea\label{CY3constraint}
\begin{array}{lll}D&\equiv&P\,\hat{D}+Q\,\tilde{D}\\ \\a_2&\equiv&P\,\hat{a}_2+Q\,\tilde{a}_2\,, \end{array}
\eea
while keeping $a_1$ completely generic. Here $P,Q$ are generic polynomials of \emph{odd} degree\footnote{More precisely, we require the Poincar\'e-dual classes of the submanifolds $\{P=0\}\subset X_3$ and $\{Q=0\}\subset X_3$ be odd classes.} on $X_3$. If $\hat{D},\tilde{D}$ were constants, we could redefine things such that  for example $D$ simply coincides with $P$, but in general this is not the case. It is now easy to see that thanks to \eqref{CY3constraint} some integral (1,1)-cycles of the D7-stack are manifest in our Calabi-Yau threefold and they can be written in the ambient four-dimensional manifold as
\bea\label{FW2cycles}
C^{\pm}_{(2)}&:&\left\{\begin{array}{l}P=0\\  Q=0\\ \xi=\pm a_1\;.\end{array}\right.
\eea
As it is also the case in \cite{Braun:2011zm}, the subvarieties \eqref{FW2cycles} are complete intersections neither with the Calabi-Yau threefold equation \eqref{genericCY3}, nor with the divisor $\{D=0\}$ representing the D7-stack. Rather, they are codimension three subspaces of the ambient, which are automatically contained in $S_2$. Moreover, one is the image of the other under the orientifold map. By construction, $c_1(S_2)$ may  now integrate to an odd number over $C^{\pm}_{(2)}$. In contrast, every 2-cycle which is complete intersection with $X_3$ gives certainly an even result for the integral of $c_1(S_2)$, since $X_3$ is an hypersurface of even class ($2c_1(B_3)$) in the ambient manifold. Notice, finally, that the constraints \eqref{CY3constraint} do not introduce any singularity in $X_3$, as can be verified by computing the gradient of eq. \eqref{genericCY3} subjected to the ansatz \eqref{CY3constraint}.

As an example, let us consider the toy model discussed in ref. \cite{Collinucci:2010gz}, namely $B_3=\mathbb{P}^3$ with homogeneous coordinates $x_1,\ldots, x_4$ and hyperplane class $H$. The associated double-cover type IIB Calabi-Yau threefold is the octic hypersurface $W\mathbb{P}^4_{1,1,1,1,4}[8]$, where $\xi$ has degree four, and the hypersurface equation is given by
\begin{equation}
x_1^8 - x_2^8 + x_3^8- x_4^8-\xi^2=0\,.
\end{equation}

Let the D7-brane be located at $D=\sum_{i=1}^4 a_i x_i$. If we wanted to determine, whether or not this brane is spin, we would have to integrate its first Chern class $c_1(D)=H$ on every possible 2-cycle, and check for even/odd results. Any 2-cycle defined as a complete intersection of $D$ with another divisor will give an even answer, due to the intersection number $H^3=2$ of this CY threefold. Yet, this brane is \emph{not} spin.

In order to detect this, one must search for holomorphic curves that cannot even be written as complete intersections of two divisors with $X_3$, but as complete intersections of three divisors in the ambient fourfold.
Let us take for simplicity the curve $C$
\begin{equation}
C:  \quad \xi  = x_4^4 \quad \cap \quad x_1 = x_2  \quad \cap \quad x_3=0\,.
\end{equation}
This curve is a $\mathbb{P}^1$ that is automatically contained in $X_3$. If we now restrict the D7-brane moduli to take the form:
\begin{equation}
D = a_1 (x_1-x_2)+ a_3 x_3\,,
\end{equation}
then it will pass through $C$, as opposed to just intersecting it. Now we may integrate $c_1(D)$ over this curve, and we will find that 
\begin{equation}
\int_C c_1(D) = 1\,.
\end{equation}

Although we have `forced' this curve upon the D7-brane by constraining the moduli, it is always present in the divisor's homology, just not holomorphically.

\subsection{F-theory lift}\label{SpFtheory}

\subsubsection{General idea}

The discussion above suggests the form of the 4-cycles on which we could detect the odd-ness of the second Chern class of the Calabi-Yau fourfold in an $Sp(N)$-singular F-theory configuration. We define the elliptically fibered Calabi-Yau fourfold with an I$_{2N}^{ns}$-singularity on  $\{D=0\}\subset B_3$ as the hypersurface
\bea\label{singularZ4}
Z^{Sp(N)}_4&:&Y^2+a_1XYZ+a_{3,N}D^N YZ^3=X^3+a_2X^2Z^2+a_{4,N}D^N XZ^4+a_{6,2N}D^{2N}Z^6\,, \nonumber\\
\eea
in the ambient fivefold given by a $W\mathbb{P}^2_{XYZ}$-bundle over $B_3$. Here $a_{i,j}$ is a section of the line bundle $K_{B_3}^{-i}\otimes \mathcal{L}_D^{-j}$, where $ \mathcal{L}_D$  indicates the line bundle defined by the vanishing locus of its section $D$.

\begin{figure}[h!] 
  \centering
      \includegraphics[width=16.5cm]{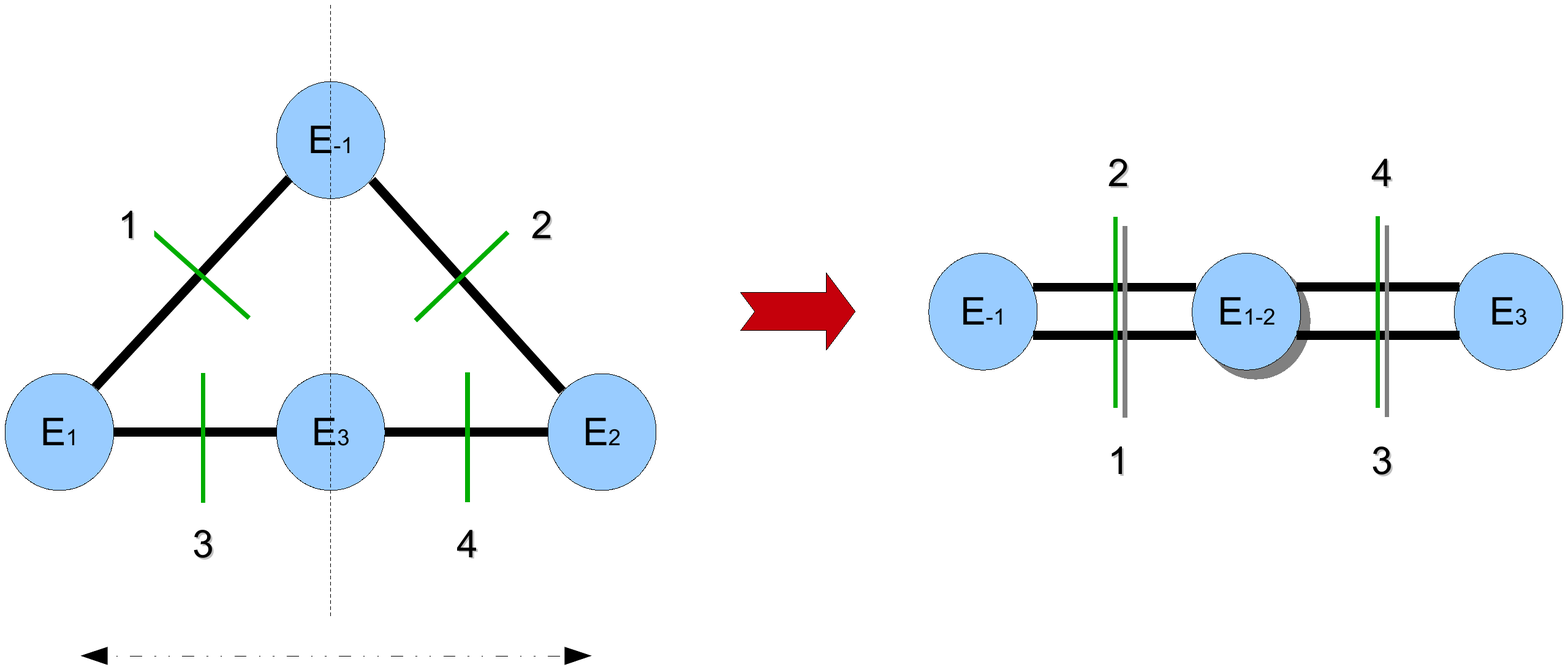}
      \caption{This shows the extended Dynkin diagram of $Sp(2)$ (right) obtained by folding the extended Dynkin diagram of $SU(4)$ (left). The monodromy responsible for the folding acts as a reflection with respect to the vertical dashed line drawn on the left figure.} \label{MonodFig}
  \end{figure}

In order to compute any topological quantity, we have to completely resolve the Calabi-Yau fourfold. The resolution procedure, as prescribed by \cite{Bershadsky:1996nh}, leads via a series of blow-ups to a resolved fiber over the singular locus which exactly reproduces the extended (or affine) Dynkin diagram of the gauge group corresponding to the singularity one starts with. For non-split singularities, like the ones in the $Sp$-series, one has to take into account non-trivial monodromies acting on the various components of the resolved fiber as we go around non-contractible paths of the singular locus. The easiest such situation, with only two exceptional divisors, is depicted in fig. \ref{MonodFig}, where $E_1$ and $E_2$ are the two components of the first exceptional divisor and they get exchanged by monodromy, while the extended node $E_{-1}$ and the last exceptional divisor $E_3$ are left invariant.
The green lines in fig. \ref{MonodFig} represent D6-branes in the type IIA picture. The resolved regime corresponds indeed to the Coulomb branch of the gauge theory living on the worldvolume of D6-branes. The latter are separated along the so called T-duality circle, i.e. one of the two fiber directions of the elliptic fibration, indicated in the figure by solid black lines. From this perspective, one realizes that the nodes of the diagram are nothing but \emph{closed} M2 branes lifting the type IIA \emph{open} strings stretched between two neighboring D6-branes. The additional direction is the so called M-theory circle, i.e. the other fiber direction of the elliptic fibration, which collapses on the D6-brane locations. Hence, in the non-split case, D6-branes as well undergo monodromies, in a way analogous to an orientifold involution. In fact, they are pairwise exchanged so to form couples of brane-image-brane ($1$ with $2$ and $3$ with $4$ in the figure). Therefore, the left part of the figure displays the extended diagram of $SU(4)$ with the usual bunch of four (separated) D6-branes. If we quotient this configuration by the monodromy action, we obtain precisely the extended diagram of $Sp(2)$, with two (separated) stacks of brane-image-brane, depicted in the right figure as shadowed green lines. 

The type IIA picture outlined above helps us finding in F-theory the 4-cycles which, lifting the 2-cycles \eqref{FW2cycles}, can be used to detect the half-quantization of the $G_4$ flux and thus the M-theory lift of the Freed-Witten anomaly. The general reasoning goes as follows.

The way we typically detect Freed-Witten anomalies \cite{Freed:1999vc} of D-branes wrapping holomorphic submanifolds  is by considering a loop of open strings attached to a given D-brane and by integrating the first Chern class of the brane worldvolume over the 2-cycle generated by the loop of string boundaries. If the result is odd, we must switch on a half-quantized gauge flux on the D-brane to cancel the anomaly. However, if such a 2-cycle is taken to be holomorphic it must degenerate to a point which is the calibrated representative of the trivial homology class of the target space. Therefore, in order to measure Freed-Witten anomalies in this case, we are forced to consider two different D-branes, like the parallel D6s in fig. \ref{MonodFig}, and look at loops of open strings stretched between them. These loops will then form 3-chains with cylindrical shape having as boundaries holomorphic 2-cycles on each of the two D-branes. Analogously to the nodes of the Dynkin diagram discussed above, these 3-chains lift in M-theory to 4-cycles, regarded as loops of the closed M2-branes lifting the type IIA open strings considered. 

Therefore, we are led to analyze the 4-cycles made by nodes of the Dynkin diagram fibered over the curve of the brane which we used to measure the Freed-Witten anomaly in type II string theory (see sec. \ref{SenLimitFW}). However, not every node works well to this end. In fact, any of the nodes depicted in the left part of fig.   \ref{MonodFig} gives rise to a 4-cycle on which the second Chern class of the fourfold integrates to an even number. One can explicitly verify this statement using the details of the geometry discussed in the next subsection. But it is instructive to realize this  from the type IIA point of view. Indeed, for instance, the integration of $c_2/2$ on the $E_1$-fibration over the curve $\{P=Q=0\}\subset B_3$ is equivalent modulo integers to the type IIA expression
\bea\label{IIAF13}
\int_{\{P=Q=0\}\subset {\rm D6}_1}F_1\;-\;\int_{\{P=Q=0\}\subset {\rm D6}_3}F_3\;,
\eea
where $F_{1,3}$ are the gauge fluxes on the D6$_{1,3}$-branes respectively and the minus sign is due to the opposite orientations of the boundaries. If we take the D6-worldvolume to be non-spin, \emph{both} $F_1$ and $F_3$ are half-integrally quantized to cancel the Freed-Witten anomaly. This means that the expression \eqref{IIAF13} is integral and thus the second Chern class of the fourfold integrates to an even number on the $E_1$-fibration. The same conclusion holds for the $E_2$-fibration (by considering the D6$_{2,4}$-branes) and in general if the fiber is any Cartan node or the extended node. To be more precise, in the cases of the last Cartan node and of the extended node, the corresponding type IIA strings stretch between one D6-brane and its image $\widetilde{\rm D6}$ under the monodromy. Hence the integration of $c_2/2$ for instance on the $E_3$-fibration in the $Sp(2)$ example would be equivalent to the integer number
\bea\label{IIAFF}
\int_{\{P=Q=0\}\subset {\rm D6}_3}F_3\;+\;\int_{\{P=Q=0\}\subset \widetilde{\rm D6}_3}{\tilde{F}_3}\;,
\eea
where the plus sign arises because the gauge flux is odd under the orientifold involution.

\begin{figure}[h!] 
  \centering
      \includegraphics[width=16.5cm]{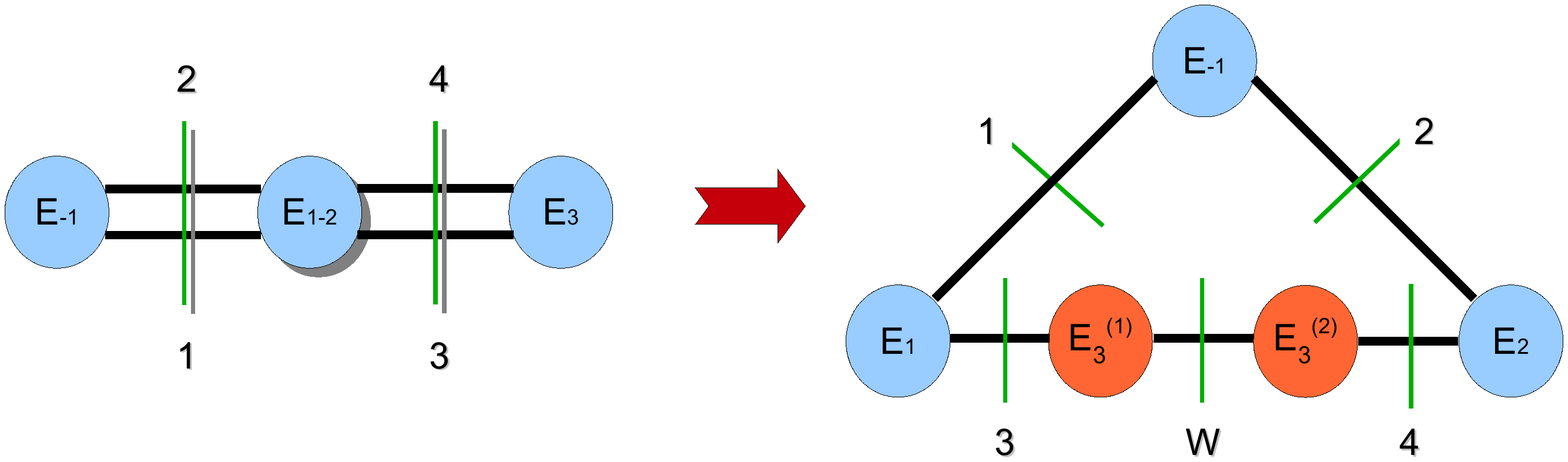} 
     \caption{This shows the transition from the extended Dynkin diagram of $Sp(2)$ (left) to the extended Dynkin diagram of $SU(5)$ (right) happening along the curve $\{P=Q=0\}$ due to the singularity enhancement. The fifth D-brane of the $SU(5)$ stack is given by the Whitney-type brane. The orange nodes are the fibers of the 4-cycles on which it is possible to detect the Freed-Witten anomaly.} \label{WhitneySp}
  \end{figure}

The above logic suggests that a node which would allow us to measure an odd value for $c_2$ could be the one which lifts an open string stretching between a D-brane of the stack and an other D-brane which carries no flux (or an integrally quantized flux). The only such D-brane around is the Whitney-type brane. This leads us to constrain the complex structure of the fourfold in such a way that our curve $\{P=Q=0\}\subset B_3$ is contained in both the D6-stack and the Whitney-umbrella brane. In other words, in order to identify the ``detecting'' 4-cycle, we have to require that the Freed-Witten 2-cycle $\{P=Q=0\}\subset B_3$ becomes a branch of the matter curve represented by the intersection between the non-abelian stack and the I$_1$-locus. Before formalizing this concept, let us explain the result in terms of diagrams. Starting from an $Sp(N)$-singularity, our constraint on the complex structure of the blown-up fourfold will be such that there will be an enhancement to $SU(2N+1)$ along the curve $\{P=Q=0\}$, as pictorially shown in fig. \ref{WhitneySp}.
As it is manifest from the figure, the only relevant change for our purpose is the splitting of the last Cartan node ($E_3$ for $Sp(2)$) into two branches ($E_3^{(1)}$ and $E_3^{(2)}$), each of which lifts the open string stretching between one D-brane of the stack and the Whitney-type D-brane W. Due to the constraint made, the shape of the Whitney-umbrella will not be completely generic and an amount of integral flux would be necessarily generated on it to keep this configuration stable. However, since we are not focusing on the dynamics of the system here, but rather we are probing it to measure possible anomalies, we can assume that no flux is present on W. Therefore, focusing on the example of fig.  \ref{WhitneySp}, the integration of $c_2/2$ on the $E_3^{(1)}$-fibration over the curve $\{P=Q=0\}\subset B_3$ is equivalent modulo integers to the type IIA expression
\bea\label{IIAF}
\int_{\{P=Q=0\}\subset {\rm D6}_3}F_3\;,
\eea
which is half-integral if and only if the stack is non-spin.  For the $E_3^{(2)}$-fibration, we find the same expression in terms of the image flux $\tilde{F}_3$.

Let us finally remark that the symmetry enhancement along the whole matter curve does not have to be $SU(2N+1)$, which only appears along the branch $\{P=Q=0\}$ due to the ansatz we will make. Indeed, for an I$_{2N}^{ns}$-Kodaira singularity, Tate's algorithm prescribes along the intersection with the I$_1$-locus an enhancement to an  I$_{2N+1}^{ns}$-Kodaira singularity, which corresponds to an unconventional gauge group \cite{Bershadsky:1996nh}. As it will be clearer later, the reason for this slightly stronger requirement is to have a closed algebraic expression for the two branches of the last Cartan node. In contrast, by just asking 
\bea\label{weaker}
W=\hat{W}P+\tilde{W}Q\,,
\eea 
where $\{W=0\}$ defines the Whitney-type D-brane W, we would still split the last Cartan node, but non-trivial monodromies would mix the two branches and we could not single out one of them globally \cite{Braun:2011zm}.

\subsubsection{Formalization} \label{formalization}

Let us now put onto rigorous ground what we have previously discussed. In order to blow-up the $Sp(N)$-singularity of \eqref{singularZ4}, we will follow the toric procedure adopted in \cite{Collinucci:2010gz}. First of all we introduce a new homogeneous coordinate $\sigma$ to the ambient space and impose the additional equation $\sigma=D$. Then we add $N$ toric vectors $v_i$ (together with $N$ projective relations), which, in order to stick with the notations of \cite{Bershadsky:1996nh}, we will label with odd numbers only\footnote{The reason is to distinguish them from the even-labeled ones which are further introduced to resolve split singularities, like the ones leading to unitary gauge groups (see sec. \ref{SUFtheory}).}, i.e. $1,\ldots,2N-1$. 

The Calabi-Yau fourfold will be defined as a complete intersection of two equations in an ambient sixfold $X_6$, given by a projective bundle over $B_3$. The ambient fiber over $B_3$ will be defined by the following assignment of projective weights
\bea\label{toric6fold}
W\mathbb{P}^3_{\sigma XYZv_1v_3\ldots v_{2N-3}v_{2N-1}}&:&\begin{array}{ccccccccc}\sigma&X&Y&Z&v_1&v_3&\cdots&v_{2N-3}&v_{2N-1}\\ \hline 0&2&3&1&0&0&\cdots&0&0\\ 1&1&1&0&-1&0&\cdots&0&0\\ 1&2&2&0&0&-1&\cdots&0&0\\ \vdots&\vdots&\vdots&\vdots&\vdots&\vdots&\ddots&\vdots&\vdots\\ 1&N-1&N-1&0&0&0&\cdots&-1&0\\ 1&N&N&0&0&0&\cdots&0&-1 \end{array}
\eea
The rows represent independent $\mathbb{C}^*$-actions on the coordinates. In order to define the ambient space, we mod these out, much like in projective spaces. The first row indicates the pre-existing projective identification of $W\mathbb{P}^2_{XYZ}$. 

Projective spaces are defined as $\mathbb{C}^n-\{ 0,\ldots,0 \} /\mathbb{C}^*$, i.e. before modding out by the action, we must first delete the origin. In toric language, one would say that the \emph{Stanley-Reisner ideal} consists of the element $z_1 \ldots z_n$, meaning that these coordinates are forbidden from vanishing simultaneously.

The fiber defined above for the sixfold $X_6$ has the following SR ideal:
\bea\label{SRideal}
SR_{Sp(N)}&:&\left\{XYZ\;;\; v_{2i-1}Z|_{i=1,\ldots,N}\;;\; v_{2i-1}XY|_{i=0,\ldots,N-1}\;;\; v_{2i-1}v_{2j-1}\left|_{\begin{subarray}{l}i,j=0,\ldots,N\\ j-i\,>1\end{subarray}}\right. \right\}\,,\nonumber\\ 
\eea 
where by convention we put $v_{-1}:=\sigma$. The above is easily derived if we express the table \eqref{toric6fold} in the basis in which it is manifest that $v_{2i-1}$ induces a blow-up along the locus $X=Y=v_{2i-3}=0$, for $i=1,\ldots,N$. It is not difficult to realize that the last elements in \eqref{SRideal}, quadratic in the $v$s, arise by taking pairwise differences of the rows in table \eqref{toric6fold} expressed in the new basis. These elements are responsible for the appearance of the ``unidimensional'' structure of the extended Dynkin diagrams of the {\bf C}-series. Calling $V\equiv \prod_{i=0}^{N}v_{2i-1}^{N-i}$, which has multi-degree $(0,1,2,\dots,N-1,N)$, the blown-up Calabi-Yau fourfold is given by the following intersection in the sixfold $X_6$:
\bea\label{genericCY4}
\tilde{Z}^{Sp(N)}_4&:&\left\{\begin{array}{cc}Y\left(Y+a_1XZ+a_{3,N}VZ^3\right)=X^3 \prod_{i=0}^{N}v_{2i-1}^{i}+&\vspace{0.3cm} \\  +a_2X^2Z^2+a_{4,N} VXZ^4+a_{6,2N}V^2Z^6& 2\times(3,1,2,\dots,N-1,N) \\ \\ \prod_{i=0}^Nv_{2i-1}=D&(0,\dots,0)\;,  \end{array}\right.\nonumber\\
\eea
where on the right the multi-degrees of the two hypersurfaces are displayed. Note that, as was the case for the Calabi-Yau threefold in sec. \ref{SenLimitFW}, these degrees are even.

The second Chern class of the resolved fourfold, which determines the quantization condition of the $G_4$ flux \cite{Witten:1996md}, is conveniently split into two pieces
\bea\label{c2SpSing}
c_2\left(\tilde{Z}^{Sp(N)}_4\right)&=&c_2\left(Z_4\right)+\Delta c_2\,,
\eea
where the first part is the second Chern class of the smooth phase of the elliptically fibered fourfold
\bea\label{c2old}
c_2\left(Z_4\right)&=&c_2(B_3)+11c_1^2(B_3)-12F^2\,,
\eea
while the second depends in an easy way on the $N$ exceptional divisors arising after the resolution 
\bea\label{c2new}
\Delta c_2&=&-\sum_{i=1}^{N}\,i\,\left(7\,c_1(B_3)-i\,\mathcal{D}\right)E_{2i-1}\,.
\eea
In the expressions above $F$ indicates the class of the 0-section $Z=0$, $\mathcal{D}$ the Poincar\'e-dual class of $\{D=0\}\subset B_3$, $E_k\,:\,v_k=0$ are the blow-up classes, and we have everywhere omitted the pull-back map acting on classes of the base. It has been proven in \cite{Collinucci:2010gz} that the class \eqref{c2old} is always even. Thus we concentrate on the class \eqref{c2new}, which, if odd, leads to a ``half''-quantized $G_4$ flux. Let us remark here that eq. \eqref{c2new} provides in this context an iteration of Fulton's formula \cite{Fulton,Andreas:1999ng} for resolved manifolds after a single blow-up, thus extending the result  for $N=1$ already discussed in \cite{Collinucci:2010gz}.

Analogously to the case of the Calabi-Yau threefold \eqref{genericCY3}, a generic equation like \eqref{genericCY4} for the Calabi-Yau fourfold does not allow us to identify the 4-cycles on which the integral of $\Delta c_2$ is odd. Therefore, we implement the general logic outlined in the previous subsection by the following ansatz for some of the polynomial coefficients of \eqref{genericCY4}
\bea\label{CY4constraint}
\begin{array}{rll}
D&\equiv&P\,\hat{D}+Q\,\tilde{D}\\ \\ a_{2k+2\,,\,kN}&\equiv&P\,\hat{a}_{2k+2\,,\,kN}+Q\,\tilde{a}_{2k+2\,,\,kN}\qquad\qquad  k=0,1,2  \;.\end{array}
\eea 
This extends conditions \eqref{CY3constraint}, which are recovered here by taking $k=0$ only, and implies condition \eqref{weaker} since
\bea
W&=&a_{4,N}^2+a_1a_{3,N}a_{4,N}-a_1^2a_{6,2N}-a_2a_{3,N}^2-4a_2a_{6,2N}\,.
\eea
By looking at the structure of the various exceptional divisors $\{v_{2i-1}=0\}$ when restricted to the curve $\{P=Q=0\}$, one easily realizes that their pattern of intersection undergoes exactly the transition displayed in fig. \ref{WhitneySp}. In particular, all of them split into two well-defined branches and all the pre-existing monodromies are killed. Indeed, on $\{P=Q=0\}$ one has
\bea\label{splitEi}
E_{2i-1}&:&\left\{\begin{array}{rll}v_{2i-1}&=&0\\ Y&=&0\end{array}\right.\quad \cup\quad \left\{\begin{array}{rlc}v_{2i-1}&=&0\\ Y&=&-a_1\end{array}\right.\qquad i=1,\ldots,N-1\;,
\eea
each of which geometrically is a $\mathbb{P}^1$ with coordinates $v_{2i-3},v_{2i+1}$ fibered over the curve. However, as already stressed, it is the last exceptional divisors which interests us more. Before the transition, $E_{2N-1}$ is a quadratic $\mathbb{P}^1$ embedded in the $\mathbb{P}^2$ with coordinates $X,Y,v_{2N-3}$. But after imposing \eqref{CY4constraint}, our Calabi-Yau fourfold gains extra $(2,2)$-cycles, due to the splitting of $E_{2N-1}$, and they look like
\bea
C^{(1)}_{(4)}&:&\left\{\begin{array}{l}P=0\\  Q=0\\ v_{2N-1}=0\\ Y=0\;\;,\end{array}\right. \label{FW4cycle+}\\ C^{(2)}_{(4)}&:&\left\{\begin{array}{l}P=0\\  Q=0\\ v_{2N-1}=0 \\ Y=-\left(a_1X+a_{3,N}\,v_{2N-3}\right)\;\;,\end{array}\right.\label{FW4cycle-}
\eea
where in \eqref{FW4cycle-} we have performed the $N$ gauge-fixings $Z=v_{2i-1}=1$ for $i=0,\ldots,N-2$, so that $V=v_{2N-3}$. We are still left with one gauge freedom, which we can use to describe the geometry of this 4-cycles. A closer look shows indeed that they are $\mathbb{P}^1$-fibrations over the curve $\{Q=0\}\cap\{P=0\}$ contained in the surface where the singularity was. The fibers are $E_{2N-1}^{(1)}$ and $E_{2N-1}^{(2)}$ respectively, in the notation of the previous subsection. The last equations of \eqref{FW4cycle+} and \eqref{FW4cycle-} tell us that $X$ and $v_{2N-3}$ are exactly the coordinates of the $\mathbb{P}^1$ fibers, due to the presence of the element $v_{2N-3}XY$ in the Stanley-Reisner ideal \eqref{SRideal}. Again these 4-cycles are not complete intersection with the Calabi-Yau fourfold. Rather they are surfaces of the ambient six-fold which are automatically contained in $\tilde{Z}^{Sp(N)}_4$, namely they automatically satisfy equations \eqref{genericCY4}.
While integrating $\Delta c_2$ over any 4-cycle which is complete intersection with  \eqref{genericCY4} gives certainly an even number, due to the even-ness of the fourfold class, the new 4-cycles may in contrast be good candidates to detect the Freed-Witten anomaly of F-theory. In fact, we find the following general formula
\bea\label{integratedc2new}
\int_{C^{1,2}_{(4)}}\Delta c_2&=&\int_{B_3}\left[7\,c_1(B_3)-(2N-1)\,\mathcal{D}\right]\,\mathcal{P}\,\mathcal{Q}\nonumber\\&=&\int_{\{P=Q=0\}\subset S_2}\left[2\,(4-N)\,c_1(B_3)|_{S_2}+(2N-1)\,c_1(S_2)\right]\,,
\eea
where $\mathcal{P},\mathcal{Q}$ are the Poincar\'e-dual class of $\{P=0\},\{Q=0\}\subset B_3$ respectively, and in the last step we have used the adjunction formula. Thanks to the simple structure of $\Delta c_2$ in \eqref{c2new}, which does not contain terms quadratic in the exceptional divisors, eq. \eqref{integratedc2new} can be easily derived. To do that, one first of all simplifies the integrand using the elements $v_{2i-1}XY|_{i=0,\ldots,N-1}$ of the SR ideal \eqref{SRideal}. They imply on $\{Y=0\}$ the following relation
\bea
E_{2N-1}\,\sum_{i=1}^N i E_{2i-1}&=&-D\,\left(2c_1(B_3)+2F-\sum_{i=1}^N i E_{2i-1}\right)+2E_{2N-1}c_1(B_3)\,.
\eea
It is easy to see that now one can factorize the class of $\{Y=0\}\cap\{D=0\}$, which is half the class of the proper transform \eqref{genericCY4}. Therefore one is led to compute an integral over the blown-up elliptic fourfold, where the following formulae hold for any type of singularity  \cite{Grimm:2010ks,Grimm:2011sk}
\bea\label{IntersNumbers}
\int_{\tilde{Z}_4}E_{k}\,\cA\,\cB\,\cC&=&0\quad\qquad\qquad\qquad\qquad\qquad k=1,\dots,{\rm rank}\,\mathbb{G}\,,\nonumber\\ 
\int_{\tilde{Z}_4}E_{k}\,E_{l}\,\cA\,\cB&=&-C_{kl}\int_{B_3}\cD\,\cA\,\cB\qquad\qquad k,l=1,\dots,{\rm rank}\,\mathbb{G}\,.
\eea
Here $\cA,\cB,\cC$ are any three divisors of the base and $C_{kl}$ is the \emph{symmetric} Cartan matrix of the Lie algebra $\mathbb{G}$ corresponding to the singularity. For the $Sp(N)$ case the latter is the  $N\times N$ matrix $C_{kl}=2\left(2\delta_{kl}-\delta_{kN}-\delta_{lN}-\delta_{k+1,l}-\delta_{k,l+1}\right)$, compatibly with \eqref{SRideal}.

The remarkable fact about formula  \eqref{integratedc2new}, as the last line suggests, is that its even/odd-ness only depends on the singular locus $S_2$ being spin or not, while it is independent of whether the class $\cD$ which defines it as a subvariety of $B_3$ is even or odd. In other words, it is only sensitive to an intrinsic property of that surface and not on the details of its embedding. More formally, the modulo two reduction of the expression in square brackets in \eqref{integratedc2new} is $w_2(S_2)$, while it does not contain information about the spin properties of the normal bundle of $S_2$ in $B_3$. On the one hand, this provides, as already argued, the mechanism for Freed-Witten anomaly cancellation of the corresponding D7-brane stack. On the other hand, it says that the degree of the polynomial defining $S_2$ as a divisor of $B_3$ plays no essential role in the quantization rule of $G_4$. This is important because, a polynomial $D$ of odd degree may well lead to a spin, and thus non-anomalous, D7-stack. For this to happen, $B_3$ itself must be non-spin.

It is not difficult to convince ourselves again the assumption \eqref{CY4constraint} we have made does not generate singularities on the Calabi-Yau fourfold \eqref{genericCY4}. By computing the gradient of eq. \eqref{genericCY4} subjected to the ansatz \eqref{CY4constraint}, one can indeed rigorously verify the absence of singularities.

Finally, it is important to note that our four-cycles, although constructed as $\mathbb{P}^1$-fibrations over curves living at brane intersections, are never actually \emph{entire} matter surfaces. The way we tune our moduli force the matter curves to become reducible, and our four-cycles are fibrations over one such component. In our case, $\{P=Q=0\}$ is but one component of the whole matter curve.
Indeed, the half-integral quantization of the gauge flux, as treated in \cite{Minasian:1997mm}, is \emph{designed} to yield correct indices for bifundamental matter at brane intersections. It would therefore be contradictory to ever find a half-integral period of such a flux over a matter curve, but not over a piece of such a curve. This is consistent with the fact that \cite{Krause:2012yh} have not registered any shift in the $G_4$ quantization on matter surfaces.

\section{Detection of the Freed-Witten anomaly in F-theory: SU-series} \label{sec:suseries}

In this section we address the same problem for F-theory configurations with an I$^{s}_M$ Kodaira singularity on a generic codimension one locus of the base. We will see that things are more subtle here both in the geometric F-theory picture and in the weakly coupled string description. We will again start with the Sen limit and afterwards study the geometrical structures of these split-type singularities by distinguishing even and odd ranks.

\subsection{Sen limit}\label{SenLimitSU}

A crucial feature that distinguishes (and makes it more subtle) the weak coupling limit of F-theory configurations with unitary-type singularities from those with symplectic-type ones is that the formers lead to Calabi-Yau threefolds with conifold singularities. The origin of these singularities is in the Tate coefficient $a_2$ having for I$^{s}_M$ a single zero along the singular locus $\{D=0\}$, instead of being generic as is the case in eq. \eqref{genericCY3}. This shape of $a_2$ is actually already contained in our constrained form \eqref{CY3constraint}. In other words, to treat Sen's limit of $SU(M)$ F-theory models, we simply have to further require in the moduli space the conditions
\bea\label{SenSU}
\hat{a}_2&\equiv&\hat{D}\,a_{2,1}\nonumber\\
\tilde{a}_2&\equiv&\tilde{D}\,a_{2,1}\,.
\eea
It is easy to see that this restriction is now sufficiently drastic to generate codimension three singularities in the Calabi-Yau threefold given by the following intersection of four divisors in the ambient four-manifold
\bea\label{conifold}
{\rm conifold\;points}&:&\left\{\begin{array}{l}\xi=0\\ a_1=0\\ D=0\\  \hat{a}_2=0\;.\end{array}\right.
\eea 

Indeed, the CY hypersurface equation assumes the shape of a conifold:
\begin{equation}
\xi^2 = a_1^2+\hat a_2\,D\,.
\end{equation}

However this restriction is essential to make visible the $U(M)$ stack\footnote{We neglect here the issue of the $U(1)$ related to the center of mass and its fate in the F-theory lift.} of D7-branes and its image-stack under the orientifold map, as was elucidated in \cite{Krause:2012yh}. Indeed, the surface $S_2$ in \eqref{invariantS2} now clearly factorizes in brane plus image-brane
\bea\label{S2pm}
S_2^\pm&:&\left\{\begin{array}{l}D=0\\ \xi=\pm a_1\;.\end{array}\right.
\eea
In contrast, in the ``constrained'' non-split situation discussed in sec. \ref{SenLimitFW}, the stack was invariant and only its curve $\{P=Q=0\}$ factorized. 

Therefore we are now able to immediately identify the 2-cycles for detection of Freed-Witten anomalies without any need of further requirements on the moduli space. Given any polynomial $Q$ with the same properties as before, the 2-cycles look like 
\bea\label{FW2cyclesSU}
C^{\pm}_{(2)}&:&\left\{\begin{array}{l}D=0\\  Q=0\\ \xi=\pm a_1\;.\end{array}\right.
\eea
Moreover, notice that these Riemann surfaces do not generically intersect the conifold singularities of $X_3$, because this would require solving five equations in the ambient four-manifold. Hence it is likely that the presence of the conifold points does not affect the evaluation of the integral of $c_1(S_2)$ over $C^{\pm}_{(2)}$. Nevertheless, being the Calabi-Yau threefold singular, the computation of Chern classes itself may no longer be reliable. And on top of that, it is not clear whether in this circumstance $w_2(S_2)$ equals $w_2({\rm N}_{X_3}\,S_2)$, and which one matters for cancellation of Freed-Witten anomalies. A procedure is therefore needed to cure this singularity. To this end, one may immediately think to a couple of possibilities, namely the small resolution of the conifold points and the deformation. However \cite{Donagi:2009ra}, both are ruled out: The former because, in the absence of $B$-field, it would lead to a Calabi-Yau threefold not invariant under the orientifold involution, the latter because it would break the unitary gauge group to the parent symplectic one. A standard resolution of the conifold points is also not acceptable, because it would be non-crepant. We will not attempt here to solve this problem at weak coupling. Rather, we will focus on finding the 4-cycles in the well-behaved F-theory which are able to detect Freed-Witten anomalies and thus give the right lift of \eqref{FW2cyclesSU}.

\subsection{F-theory lift}\label{SUFtheory}

Let us now discuss in detail the structure of the F-theory fourfold in the presence of an I$^{s}_N$ Kodaira singularity and its complete resolution. Once again, our analysis aims at finding suitable integral $(2,2)$-cycles for detecting shifted quantization conditions for the $G_4$ flux. It turns out to be convenient to split the discussion in two sub-cases, namely $SU(even)$ and $SU(odd)$ gauge groups in eight dimensions. In appendix \ref{GeneralSU} we have collected general formulae holding for all $SU(M)$ singularities, which we will need throughout this section.

\subsubsection{The SU(2N) family}\label{SUevenFamily}

We start from the even case as it is easier to deal with. The reason is that the I$^{s}_{2N}$ family of Kodaira singularities is very closely related to the I$^{ns}_{2N}$. By looking at the table of the Tate algorithm in \cite{Bershadsky:1996nh}, one indeed realizes that the only difference is the one already encountered for the Sen limit in sec. \ref{SenLimitSU}, i.e. $a_2$ is generic in the non-split case, while it has a single zero along $\{D=0\}$ in the split case. This strongly suggests that we should treat I$^{s}_{2N}$ singularities in a similar manner as we did for I$^{ns}_{2N}$ in sec. \ref{SpFtheory}.

Since $SU(2)=Sp(1)$ has already been discussed in the previous section, the present analysis  concerns $SU(2N)$ gauge groups for $N\geq2$. Recall, however, that the prescription given in  \cite{Bershadsky:1996nh} for resolving I$^{s}_{2N}$ requires to add further toric coordinates  $v_{2i}$ for $i=1,\ldots,N-1$, with new projective $\mathbb{C}^*$-actions $\rho_{2i}$ such that $\sigma,X,Y, v_{2i}$ have weights $(1,i,i+1, -1)$   respectively under $\rho_{2i}$. We refer to appendix \ref{GeneralSU} for the details of the geometry.

\begin{figure}[h!] 
  \centering
      \includegraphics[width=16.5cm]{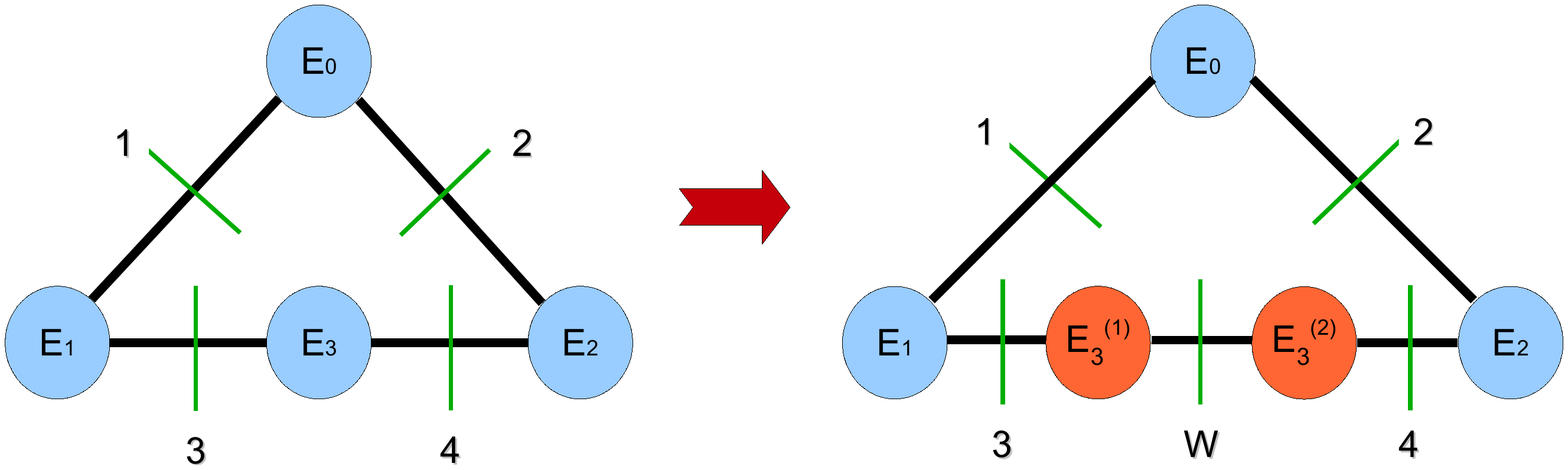} 
     \caption{This shows the transition from the extended Dynkin diagram of $SU(4)$ (left) to the extended Dynkin diagram of $SU(5)$ (right) happening along the curve $\{P=Q=0\}$ due to the singularity enhancement. The fifth D-brane of the $SU(5)$ stack is given by the Whitney-type brane. The orange nodes are the fibers of the 4-cycles on which it is possible to detect the Freed-Witten anomaly.} \label{WhitneySU}
  \end{figure}

One can now easily seek for the detecting 4-cycles in analogy with the analysis done for the $Sp(N)$ singularities in sec. \ref{SpFtheory}, by imposing that:
\begin{eqnarray}
D&\equiv&P\,\hat{D}+Q\,\tilde{D} \\ a_{2k+2\,,\,kN}&\equiv&P\,\hat{a}_{2k+2\,,\,kN}+Q\,\tilde{a}_{2k+2\,,\,kN}\qquad\qquad  k=1,2  \,,\end{eqnarray}
and $a_2 = a_{2,1}\,D$\,.
The ansatz makes the gauge symmetry enhance on $\{P=Q=0\}$ from $SU(2N)$ to $SU(2N+1)$ along the whole matter curve. The enhancement manifest itself as the splitting into two of the node $E_{2N-1} \mapsto E_{2N-1}^{(1)} \cup E_{2N-1}^{(2)}$.   Such transition is shown in fig. \ref{WhitneySU} for the $N=2$ case.

Focusing on the first component, we find a 4-cycle $C^{(1)}$ given by the fibration of $E_{2N-1}^{(1)} $ over the curve detecting the Freed-Witten anomaly in type IIB:

\bea\label{FW4cycle+SU}
C^{(1)}_{(4)}&:&\left\{\begin{array}{l}P=0\\  Q=0\\ v_{2N-1}=0\\ Y=0\;\;,\end{array}\right. 
\eea
One can prove that the general formula for the integral of $c_2(\tilde{Z}_4^{SU(M)})$, on the surface \eqref{FW4cycle+SU} is the following:
\bea \label{SUevenIntegral}
\int_{C_{(4)}^{(1)}} c_2 &=&\int_{B_3} \left( 6\,c_1(B_3)-(2\,N-1) \cdot \mathcal{D} \right) \cdot \mathcal{P}\cdot \mathcal{Q}\,.
\eea

This shows that the even or oddness of $c_2(\tilde{Z}_4^{SU(M)})$ is correlated with the spin-ness or non spin-ness of the brane stack on $\mathcal{D}$. Again, we emphasize that, as explained in the last paragraph of \ref{formalization}, this 4-cycle cannot be a matter surface. Those should always yield integral periods for $G_4$.

We will outline the arguments behind the proof, which is rather tedious in its detail: 

The general elements of the Stanley-Reisner ideal found in appendix \ref{GeneralSU} imply that, on the 4-cycle \eqref{FW4cycle+SU}, the following coordinates can be taken to be different from zero, and `gauge-fixed' to one: $\sigma, v_1, \ldots, v_{2N-3}, Z$. This means that the respective divisor classes vanish on the 4-cycle. Only $X, v_{2N-2}$ remain unfixed. 

Since our goal is to relate an integral on the ambient sixfold $X_6$ of the fourfold $\tilde{Z}_4^{SU(M)}$ to an integral on $B_3$, we must identify a form that is Poincar\'e dual to $B_3$ in $X_6$. In this case, we may choose one of the following two equivalent six-forms:
\begin{equation}
[Y] \cdot [v_{2N-1}] \cdot [X] \qquad {\rm or} \qquad [Y] \cdot [v_{2N-1}] \cdot [v_{2N-2}]\,.
\end{equation}
One can see that this works, because after setting all of these coordinates to zero,  those on $B_3$ remain unfixed. Let us pick the latter $PD_{B_3} \equiv  [Y] \cdot [v_{2N-1}] \cdot [v_{2N-2}]$.

Let us calculate the total Chern class $c(\tilde{Z}_4^{SU(M)})$ with the adjunction formula:
\begin{equation}
c(\tilde{Z}_4^{SU(M)})  = \frac{c(B_3) \cdot (1+[x]) \cdot (1+[y]) \cdot (1+[z]) \cdot (1+[\sigma]) \cdot \Pi_{i=1}^{2N-1}(1+[v_i])}{\left((1+\mathcal{D}) \cdot (1+[x]+[y]+c_1(B_3))\right)}\,.
\end{equation}
By restricting onto $C_{(4)}^{(1)}$, the SR ideal allows us to get rid of a lot of terms:
\begin{equation}
c(\tilde{Z}_4^{SU(M)})|_{C_{(4)}^{(1)}}  = \frac{c(B_3) \cdot (1+x) \cdot (1+y) \cdot (1+v_{2N-2}) \cdot (1+v_{2N-1})}{\left((1+\mathcal{D}) \cdot (1+x+y+c_1(B_3))\right)}\,.
\end{equation}
For convenience, we will denote a divisor class of a coordinate $p$ simply as $p$, and drop the $[]$.
We must now exploit all linear relations between divisor classes, in order to express $c_2(\tilde{Z}_4^{SU(M)})$ in terms of the classes that vanish, and of $v_{2N-2}$. The general relations, \emph{after restricting onto} $C_{(4)}^{(1)}$, are the following:
\begin{equation}
[v_{2N-1}] = \mathcal{D}-[v_{2N-2}]\,; \quad [X] = 2\,c_1(B_3)-N\,\mathcal{D}+[v_{2N-2}]\,; \quad [Y] = 3\,c_1(B_3)-N\,\mathcal{D}\,.
\end{equation}

Dropping terms with four indices along $B_3$, we can then extract the second Chern class:
\begin{equation}
c_2(\tilde{Z}_4^{SU(M)})|_{C_{(4)}^{(1)}} =(4\,c_1(B_3)-2\,\mathcal{D}) \cdot [v_{2N-2}]-[v_{2N-2}]^2\,.
\end{equation}
Finally, from the relation in the SR ideal $[v_{2N-2}] \cdot [X]=0$, we find that $[v_{2N-2}]^2=(N\,\mathcal{D}-2\,c_1(B_3)) \cdot [v_{2N-2}]$. From this, we arrive at our general formula \eqref{SUevenIntegral}.

\subsubsection{The SU(2N+1) family}\label{SUoddFamily}

We are left to address the odd series of unitary groups, which is more subtle. We will actually develop a method for detecting the shift in quantization of $c_2$ for \emph{all} unitary groups with $SU(M \geq 5)$, also including the even ones. The case $SU(3)$ is special, and is remanded to appendix \ref{SU(3)}

By looking at the Tate table \cite{Bershadsky:1996nh}, we see that in order to enforce an I$^{s}_{2N+1}$ singularity on top of an existing I$^{s}_{2N}$  two more polynomials have to vanish at linear order on $\{D=0\}$, i.e. $a_{4,N}$ and $a_{6,2N}$. Therefore, the ansatz for \emph{all} the $a$-coefficients in \eqref{CY4constraint} becomes automatically satisfied.

\begin{figure}[h!] 
  \centering
      \includegraphics[width=16.5cm, height=12cm]{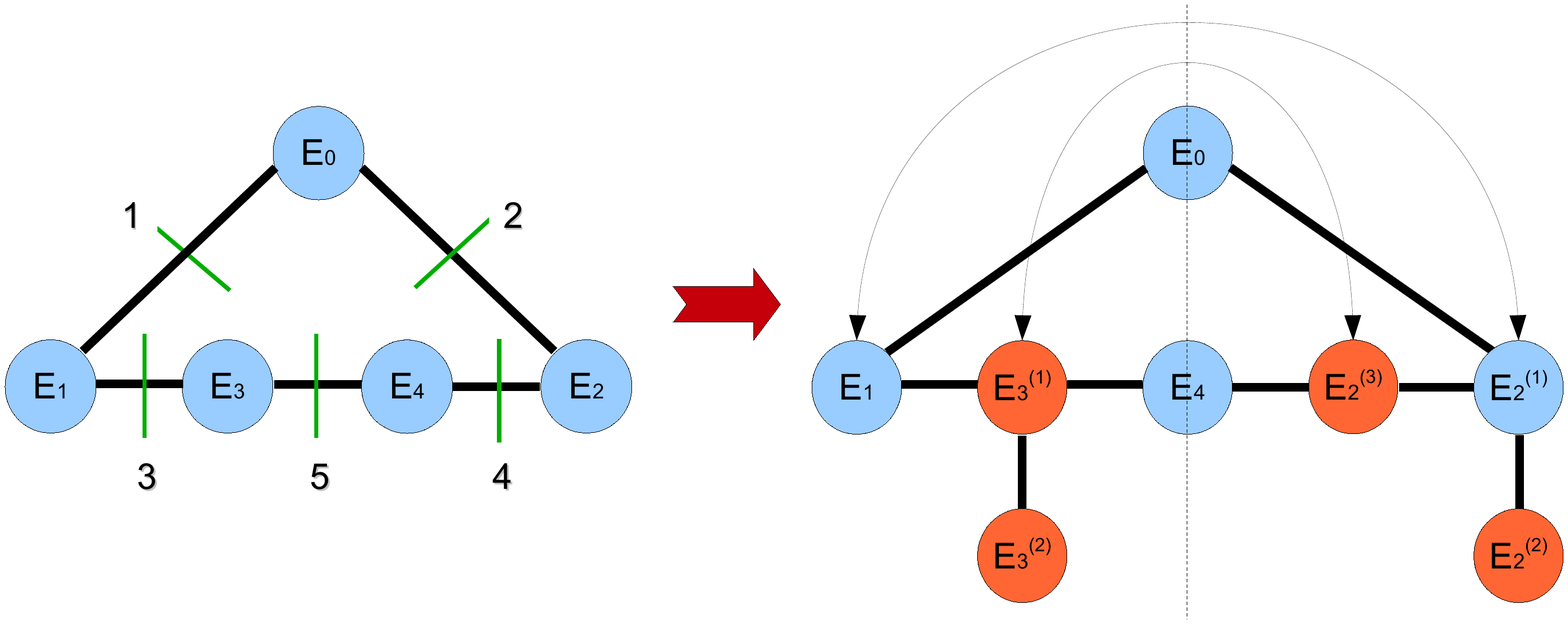} 
     \caption{This shows the transition from the extended Dynkin diagram of $SU(5)$ (left) to the extended Dynkin diagram of $SO(10)$ (right) happening along the curve $\{P=Q=0\}$ due to the singularity enhancement. Nodes connected by arrows are identified. The orange nodes are the fibers of the 4-cycles on which it is possible to detect the Freed-Witten anomaly.} \label{OrientSU}
  \end{figure}

A type IIA argument totally analogous to the ones given in sec. \ref{SpFtheory} tells us that we can no longer use the branches of the last exceptional divisor to identify the detecting 4-cycles. Indeed, they now lift open strings stretching between two ordinary branes of the $SU(2N+1)$-stack, both of them carrying an half-quantized gauge flux in the non-spin case. Therefore we have to look elsewhere. The only other matter curve available is the intersection of the non-abelian stack and the orientifold plane, i.e. $\{D=a_1=0\}\subset B_3$. Like the Whitney-type D-brane before, the orientifold has the property of having no gauge flux on its worldvolume. But it also does not admit ending open strings. However, we could consider the \emph{unoriented} open strings, which start from a given D-brane of the stack, loop around the O-plane and come back to the D-brane with the opposite orientation. If we now consider a loop of such open string worldsheets, this will no longer be a 3-chain, but rather a 3-cycle. The latter, in turn, will intersect the D-brane on the 2-cycle originating from gluing together the loops of the two oppositely-oriented string boundaries. Finally, this 3-cycle will again lift in M-theory to the 4-cycle we are looking for. 

This argument suggests that we should constrain the complex structure of the blown-up fourfold, which is given in eq. \eqref{CY4SU}, in such a way that the curve $\{P=Q=0\}\subset B_3$ is automatically contained in both the D-brane stack and the orientifold plane. Therefore, we impose the following conditions
\bea\label{CY4constraintOrient}
\begin{array}{lll}D&\equiv&P\,\hat{D}+Q\,\tilde{D}\\ \\a_1&\equiv&P\,\hat{a}_1+Q\,\tilde{a}_1\,, \end{array}
\eea
since the polynomial defining the O7-plane is
\bea
{\rm O7}&:&h\,=\,a_1^2+4a_{2,1}D\,.
\eea
Since the curve $\{P=Q=0\}$ is a branch of the intersection between the non-abelian stack and the O-plane, we experience on it the gauge symmetry enhancement from $SU(2N+1)$ to $SO(4N+2)$. This transition is pictorially shown for the $N=2$ case in fig. \ref{OrientSU} and it works the same way for $SU(2N)$ too, which enhances to $SO(4N)$ . Hence this provides an additional, alternative way of detecting Freed-Witten anomalies for $SU(2N)$ F-theory configurations. The only notable exceptions are $SU(2)$, which enhances to the III-Kodaira singularity and $SU(3)$, which enhances to the IV$^s$-Kodaira singularity \cite{Cvetic:2010rq}. Since $SU(2)=Sp(1)$, we have already described for it the rules for the quantization of the $G_4$ flux. In contrast, we have not been able to detect the Freed-Witten anomaly for $SU(3)$ F-theory configurations. In fact, we argue that for those configurations the second Chern class of the fourfold is \emph{always} even and in appendix \ref{SU(3)} we give an argument  in favor of this conjecture. Nevertheless, it is not yet clear which is the mechanism for Freed-Witten anomaly cancellation in their weak coupling limit if the $SU(3)$-stack is non-spin but no half-quantized gauge flux is induced. We hope to come back to this issue in the near future. 

The pattern of symmetry enhancement of $SU(M)$ ($M\geq4$) to orthogonal symmetries on the O-plane can be described as follows, at the level of  Dynkin diagrams (see fig. \ref{OrientSU}). The Cartan node of the exceptional divisor $E_2$ splits into three branches, one of which gets identified with $E_1$ and therefore cannot be used for detecting Freed-Witten anomalies\footnote{Recall that $E_1$ interpolates between two ordinary D-branes of the stack. }, while the other two can be used as fibers for two good detecting 4-cycles\footnote{There is a subtlety here. For $SU(4)$ only this is not a priori clear because of the extra requirement of factorization one has to impose for the $SO(8)$ enhancement  \cite{Bershadsky:1996nh}. Therefore our second method works correctly starting from $SU(5)$.}. Moreover, the Cartan nodes of $E_{i}$ for $i=3,\ldots,M-2$ separately split into two branches, all of which can in principle be used as fibers for two good detecting 4-cycles. Each of those gets identified with one branch belonging to each of the two neighboring nodes, with the only exception of one branch of $E_{M-2}$. The latter, together with one branch of $E_2$, $E_0$ and the last divisor $E_{M-1}$, does not undergo any identification. Thus, in total, there are $M-3$ identifications. One can easily see that this pattern exactly reproduces the extended Dynkin diagram of the $SO(2M)$ gauge group.  

All of this can be easily verified from the general algebraic equation of the Calabi-Yau hypersurface given in appendix \ref{GeneralSU}. In particular, the explicit expressions of some of these new nodes fibered over the curve $\{P=Q=0\}$ are particularly easy. They are the branches which do undergo identifications
\bea\label{FW4cycle+SUO}
C_{(4)j}^{(1)}&:&\left\{\begin{array}{l}P=0\\  Q=0\\ v_{j-1}=0\\ v_{j}=0\;\;,\end{array}\right.\qquad\qquad j=2,\ldots,M-2\;.
\eea 
As stressed above, the first of these new 4-cycles ($j=2$) is not relevant for our purposes of detecting anomalies. The general formula of the integral of $c_2(\tilde{Z}_4)$, given in eq. \eqref{c2SUgroups}, on the surfaces \eqref{FW4cycle+SUO} with, say, $j=3$ is given by the following very simple expression:
\bea
\int_{C_{(4)3}^{(1)}}c_2(\tilde{Z}_4)&=&\int_{B_3} \mathcal{D} \cdot \mathcal{P} \cdot \mathcal{Q}\,.
\eea
The proof proceeds very much analogously to the case covered in the previous section:

One realizes that, on $C_{(4)3}^{(1)}$, the following coordinates are fixed to be non-zero: \\
$(X, Y, \sigma, Z, v_5, \ldots, v_{M-1})$. Hence, their divisor classes vanish on the 4-cycle. We then identify the following six-form as the Poincar\'e dual to $B_3$ in the ambient sixfold $X_6$:
\begin{equation}
PD_{B_3} \equiv v_1 \cdot v_2 \cdot v_3\,.
\end{equation}

We will also need the following linear relations, valid on $C_{(4)3}^{(1)}$:
\begin{equation}
[v_2] =-2\,c_1(B_3)+2\,\mathcal{D}-[v_1]\,; \quad [v_3] = -c_1(B_3)+\mathcal{D}-[v_1]\,; \quad [v_4] = 3\,c_1(B_3)-2\,\mathcal{D}+[v_1]\,.
\end{equation}
Finally, we need the relation $[v_1] \cdot [v_4]=0$, which gives us $[v_1]^2 = [v_1] \cdot (2\,\mathcal{D}-3\,c_1(B_3))$

Now, we substitute all of these relations into the adjunction formula for the Chern class of $\tilde{Z}_4$, and arrive at the following result:
\begin{eqnarray}
\int_{C_{(4)3}^{(1)}} c_2(\tilde{Z}_4)  &=& \int_{X_6} [v_1] \cdot [v_2] \cdot [v_3] \cdot \mathcal{D} \cdot \mathcal{P}\cdot \mathcal{Q}\,.\\
&=& \int_{B_3} \mathcal{D} \cdot \mathcal{P}\cdot \mathcal{Q} = \int_\mathcal{C} \mathcal{D}\,.
\end{eqnarray}

Again, we emphasize that, as explained in the last paragraph of \ref{formalization}, this 4-cycle cannot be a matter surface. Those should always yield integral periods for $G_4$.

\subsection{The example of SU(5)}

We will now illustrate all of our general formulae in the interesting case of $SU(5)$. The Weierstrass model for the fourfold such that it creates an $SU(5)$ singularity along a divisor given by $D=0$ in $B_3$ is given by the following complete intersection:
\begin{eqnarray}
&&Y^2+a_1\,X\,Y\,Z+a_{3,2}\,\sigma^2\,Y\,Z^3 = \\ \nonumber
&&\qquad X^3+a_{2,1}\,\sigma\,X^2\,Z^2+a_{4,3}\sigma^3\,X\,Z^4+a_{6,5}\,\sigma^5\,Z^6\\
&&D=\sigma\,,
\end{eqnarray}
where $\sigma$ is an auxiliary coordinate. This space is singular at $X=Y=\sigma=0$. In order to resolve it, we introduce four new coordinates $v_1, \ldots, v_4$. The full ambient space of the resolved fourfold $\tilde Z_4$ is a projective bundle over $B_3$. The fiber is a toric space described by the following table:

\bea\label{toric6foldSU5}
W\mathbb{P}^3_{\sigma XYZv_1\ldots v_{4}}&:&\begin{array}{cccccccc}\sigma&X&Y&Z&v_1&v_2&v_3&v_{4}\\ \hline 0&2&3&1&0&0&0&0\\ 1&1&1&0&-1&0&0&0\\ 0&0&1&0&1&-1&0&0\\ 0&1&0&0&0&1&-1&0 \\ 0&0&1&0&0&0&1&-1 \end{array}
\eea

The SR ideal for this space is the following:
\bea
SR_{SU(5)}&:&\Big\{XYZ\;;\; \sigma X Y\;;\; v_iZ|_{i=1,\ldots,4}\;;\; v_{1}Y\;; v_3 Y\;; v_2 X\;; \sigma v_3\;;\sigma v_4\;; v_1 v_4 \Big\}\,.
\eea

The resolved fourfold $\tilde Z_4$ is then given by the following complete intersection:
\bea\label{CY4SU5}
\tilde{Z}^{SU(M)}_4&:&\left\{\begin{array}{l}v_{2} v_4Y^2+a_1XYZ+a_{3,2}\sigma^2 v_1 v_2 YZ^3=\vspace{0.3cm} \\ =X^3 v_1 v_3^2 v_4+ a_{2,1} \sigma v_1 v_3 X^2Z^2+\vspace{0.3cm}  +a_{4,3}\sigma^3 v_1^2 v_2 v_3 XZ^4+a_{6,5} \sigma^5 v_1^3 v_2^2 v_3 Z^6 \\ \\ \\
\sigma v_1 v_2 v_3 v_4=D\;.  \end{array}\right. 
\eea

We now impose the following ansatz on the complex structure moduli of our resolved fourfold:
\bea
\begin{array}{lll}D&\equiv&P\,\hat{D}+Q\,\tilde{D}\\ \\a_1&\equiv&P\,\hat{a}_1+Q\,\tilde{a}_1\,, \end{array}\,,
\eea
where $P$ and $Q$ are polynomials in $B_3$ defining a holomorphic curve $\mathcal{C}$ as the complete intersection in $B_3$. The first line of this ansatz simply forces that $\mathcal{C} \subset \tilde Z_4$ to be contained on the 7-brane at $D=0$, whereas the second enforces that $a_1$ vanish along it. This is the putative curve along which one measures the FW shift in the gauge flux quantization.

Now, we construct our distinguished 4-cycle as a complete intersection of four equations in $X_6$:
\begin{equation}
C_{(4)}: \quad P=0\quad \cap\quad Q=0 \quad\cap\quad v_2= 0\quad \cap \quad v_3 =0\,.
\end{equation}
This is the 4-cycle along which we will detect the shift in the quantization of $G_4$.

We note, that on $C_4$, the following coordinates are barred from vanishing:
$Z, X, Y, \sigma$, leaving only $v_1$ and $v_4$ unfixed. 

By following the procedure described in \ref{SUoddFamily}
By applying the linear relations, we arrive at the following result:
\begin{eqnarray}
\int_{C_{(4)}} c_2(\tilde{Z}_4)  &=& \int_{X_6} [v_1] \cdot [v_2] \cdot [v_3] \cdot \mathcal{D} \cdot \mathcal{P}\cdot \mathcal{Q}\,.\\
&=& \int_{\mathcal{C}} \mathcal{D}\,.
\end{eqnarray}
In the last line, we use the fact that, once we set $v_1=v_2=v_3=0$, then all coordinates of $X_6$ are `gauge-fixed' to one, except for those of $B_3$. Hence, this locus corresponds to an embedded copy of $B_3$ in $X_6$.

This formula should be interpreted as follows: Suppose we have a IIB setup with a stack of branes along a divisor $\{D=0\}$, that we suspect is a non-spin four-dimensional manifold. Suppose that its non-spin-ness can be detected on some curve $\mathcal{C}$ given by $\{P=Q=0\} \subset B_3$. In other words, suppose that
$\int_{\mathcal{C}} D $ is odd\footnote{This assumes that $B_3$ is itself spin, i.e. $c_1(B_3)$ is even. We hope to address the case of non-spin $B_3$ in the future.}. Then $c_2(\tilde{Z}_4)$ will be odd along the so-constructed complex surface $C_{(4)}$ that is a $\mathbb{P}^1$-fibration over $\mathcal{C}$.
Again, we emphasize that, as explained in the last paragraph of \ref{formalization}, this 4-cycle cannot be a matter surface. Those should always yield integral periods for $G_4$.

\section{U(1)-restrictions} \label{sec:u1restrictions}

Having treated situations where the Calabi-Yau fourfold develops symplectic and unitary singularities along a divisor on the base manifold, we would now like to take our analysis one step further, by implementing yet another restriction on the complex structure moduli space: The so-called \emph{$U(1)-restriction$}. This restriction, introduced in \cite{Grimm:2010ez}, consists in setting the $a_6$ polynomial in the `Tate form' of the Weierstrass equation to zero: $a_6 \equiv 0$. 
The effect of this restriction can be understood from the perturbative IIB point of view as follows: The setups we have been considering contain a stack of 7-branes with $U(N)$ or $Sp(N)$ gauge groups along some divisor $\mathcal{D}$. However, the tadpole cancellation of the axio-dilaton implies that there must be another 7-brane that saturates the rest of the available negative charge generated by the O7-plane. This brane is generically invariant under the orientifold involution, and hence carries gauge group $O(1)$. 

Imposing the $U(1)$ restriction splits up this remaining brane into a brane/image-brane pair, thereby allowing the pair to enhance to a $U(1)$ gauge group. This situation is more interesting if one wishes to be able to turn on chirality-inducing fluxes.

From the F-theory point of view, a singularity is created along the curve where the brane meets its image. The singularity type is recognizable as a curve of conifolds. To be more explicit, let us take again our example of the $SU(5)$ model. The $U(1)$-restricted Tate form takes the form
\begin{eqnarray}
&&v_{2} v_4Y^2+a_1XYZ+a_{3,2}\sigma^2 v_1 v_2 YZ^3 
=X^3 v_1 v_3^2 v_4+ a_{2,1} \sigma v_1 v_3 X^2Z^2+\vspace{0.3cm}  +a_{4,3}\sigma^3 v_1^2 v_2 v_3 XZ^4 \nonumber
\end{eqnarray}
which can be rewritten as follows:
\begin{equation}
Y\, \tilde Y = X \, K\,,
\end{equation}
where
\begin{eqnarray}
\tilde Y &\equiv& v_{2} v_4Y+a_1XZ+a_{3,2}\sigma^2 v_1 v_2 Z^3 \,, \nonumber \\
K &\equiv& X^2 v_1 v_3^2 v_4+ a_{2,1} \sigma v_1 v_3 XZ^2+\vspace{0.3cm}  +a_{4,3}\sigma^3 v_1^2 v_2 v_3 Z^4\,.
\end{eqnarray}
This is manifestly singular at the locus $Y=\tilde Y = X = K = 0$. In order to resolve this, we can apply the method of the small resolution, as was done in \cite{Braun:2011zm}. We construct a $\mathbb{P}^1$-bundle over $X_6$ with coordinates $(s, t)$, and define the resolved fourfold as the solution space of the matrix equation:
\[ \left( \begin{array}{cc}
Y & K  \\
X & \tilde Y 
\end{array} \right) \left( \begin{array}{c}
s \\
t  
\end{array} \right) =0\,, \]
intersected with the usual equation
\begin{equation}
D = \sigma v_1 \ldots v_{4}\,.
\end{equation}

We could also pick the transposed equation, but for simplicity we will choose this one. In order for this equation to be consistent, the new coordinates must transform appropriately under the $\mathbb{C}^*$-actions that act on all other coordinates. It is sufficient to impose the following linear relations among divisor classes: 
\begin{equation}
[s]-[t] = [K]-[Y] = [\tilde Y]-[X]\,.
\end{equation}
An additional $\mathbb{C}^*$-action is introduced such that $(s,t)$ form a $\mathbb{P}^1$. The new ambient space, which is a $\mathbb{P}^1$-bundle over $X_6$, is a \emph{seven-fold} $X_7$. Its Stanley-Reisner ideal $SR_{X_7}$ is simply given by all elements of $SR_{X_6}$, plus the element $s t$.

Let us study the cases $SU(M \geq 5)$ for simplicity.  The 4-cycles that we defined in the previous section for detecting shifted quantization conditions are now defined by the same equations as before, supplemented by the equation $s=0$:
\bea
C_{(4)} &:& P=0 \quad \cap \quad Q=0\quad \cap \quad  v_2=0 \quad \cap \quad v_3=0 \quad \cap \quad s=0\,.
\eea
Now, we apply  similar techniques as before to carry out our calculations, and arrive at the following general formula:

\begin{eqnarray}
\int_{C_{(4)}} c_2(\tilde{Z}_4)  &=& \int_{X_7} [v_1] \cdot [v_2] \cdot [v_3] \cdot [s] \cdot \mathcal{D} \cdot \mathcal{P}\cdot \mathcal{Q}\,.\\
&=& \int_{B_3} \mathcal{D} \cdot \mathcal{P}\cdot \mathcal{Q}\,.
\end{eqnarray}

Hence, the $U(1)$-restriction does not alter the results on quantization for $SU(M \geq 5)$. For the $Sp$-series we conjecture that odd-rank $Sp$ groups lead to even second Chern classes and vice-versa.
This is because in the odd-rank cases the two branches of the Whitney umbrella brane are separately non-spin and the induced flux on them is also half-integral. This effect cancels the analog effect arising on the non-abelian stack. In contrast, in the even-rank cases, the branches of the Whitney umbrella are spin and the induced gauge flux is integrally quantized. For this reason we still find, by explicit computation, an odd second Chern class also for the $U(1)$-restricted $SU(4)$-model.

\section{Outlook} \label{outlook}
Our analysis suggests many directions for further investigation. 
For some gauge groups, we find that it is the (non)-spin-ness of the tangent bundle of the D7-stack that will determine, whether $G_4$ in the corresponding resolved fourfold will be integrally quantized or not.
In others, it is the normal bundle in $B_3$ that matters. For even $c_1(B_3)$, both have the same implications. The case for non-spin base manifold needs clarification.

The Sen limit for $SU(N)$ gauge groups induces conifold singularities in the CY double-cover of $B_3$ that do not seem to admit any viable resolutions. In order to decide on topological matters such as the spin-ness of a divisor, we need to know the appropriate treatment of such singularities.

The elusive case of $SU(3)$ needs elucidation. Its peculiarity probably stems from the fact that its enhancement along the orientifold does not yield an $SO(6)$ group.

\subsection*{Acknowledgements}
A. C. is a Research Associate of the Fonds de la Recherche Scientifique F.N.R.S. (Belgium). R. S. wants to thank the Theory Division of CERN for hospitality at some stage of this work.
We would like to thank A. Braun, I. Garcia-Etxebarr\'ia, T. Grimm, W. Lerche for discussions and expecially R. Valandro for initial collaboration to the project.

\appendix

\section{Geometry of SU(N) F-theory configurations}\label{GeneralSU}

We collect here some general formulae valid for the blown-up elliptic fourfold describing an $SU(M)$ F-theory configuration. Let us define for convenience the operations $\bar{x}:=x\,{\rm mod}2$, $\hat{x}:=[(x-1)/2]$, with $\hat 0 = 0$, and $\tilde{x}:=[x/2]$, where $[\cdot]$ extracts the integer part.

Our Calabi-Yau fourfolds $\tilde{Z}^{SU(M)}_4$ are defined as complete intersections in an ambient sixfold $X_6$, constructed as projective bundles over $B_3$.
The table of projective weights defining the ambient fiber over $B_3$, in a slightly different basis with respect to \eqref{toric6fold} looks like
\bea\label{toric6foldSU}
W\mathbb{P}^3_{\sigma XYZv_1v_2\ldots v_{M-2}v_{M-1}}&:&\begin{array}{cccccccccc}\sigma&X&Y&Z&v_1&v_2&v_3&\cdots&v_{M-2}&v_{M-1}\\ \hline 0&2&3&1&0&0&0&\cdots&0&0\\ 1&1&1&0&-1&0&0&\cdots&0&0\\ 0&0&1&0&1&-1&0&\cdots&0&0\\ 0&1&0&0&0&1&-1&\cdots&0&0\\ \vdots&\vdots&\vdots&\vdots&\vdots&\vdots&\vdots&\ddots&\vdots&\vdots\\ 0&\overline{(M-1)}&\overline{M}&0&0&0&0&\cdots&1&-1 \end{array}
\eea
From the table \eqref{toric6foldSU} one can easily infer the Stanley-Reisner ideal of the ambient variety. However, one encounters a subtlety for $SU(3)$, which has to be treated separately. Therefore, for $M\neq3$, the SR ideal is made of $2+M(M-1)/2$ elements and it looks like  
\bea
SR_{SU(M)}&:&\Big\{XYZ\;;\; v_0XY\;;\; v_iZ|_{i=1,\ldots,M-1}\;;\; v_{2i-1}Y|_{i=1,\ldots,\widehat{M}}\;;\;\nonumber \\  && \quad v_{2i}X|_{i=1,\ldots,\widetilde{M}-1}\;;\; v_iv_j\left|_{\begin{subarray}{l}i,j=0,\ldots,M-1\\ j-i\,>2\end{subarray}}\right. \Big\}\,,
\eea
where we have defined $v_0:=\sigma$. 

The Stanley-Reisner ideal for $SU(3)$ has instead $6$ elements
\bea\label{SRSU3}
SR_{SU(3)}&:&\Big\{XYZ\;;\; v_0XY\;;\; v_1Z\;;\; v_2Z\;;\; v_1Y\;;\; v_0Xv_2  \Big\}\,.
\eea

The general formula for the proper transform, valid for every $M$, is
\bea\label{CY4SU}
\tilde{Z}^{SU(M)}_4&:&\left\{\begin{array}{l}\prod_{i=1}^{\widehat{M}}v_{2i}Y^2+a_1XYZ+a_{3,\widetilde{M}}v_0\prod_{i=0}^{M-1}v_i^{\widetilde{M}-\hat{i}-1}YZ^3=\vspace{0.3cm} \\ =X^3 \prod_{i=1}^{\widetilde{M}}v_{2i-1}^i\prod_{j=1}^{\widehat{M}}v_{2j}^{j-1}+ a_{2,1}v_0\prod_{i=1}^{\widetilde{M}}v_{2i-1} X^2Z^2+\vspace{0.3cm} \\ +a_{4,\widehat{M}+1}v_0\prod_{i=0}^{M-1}v_i^{\widehat{M}-\tilde{i}} XZ^4+a_{6,M}v_0 \prod_{i=0}^{M-1}v_i^{M-i-1}Z^6 \\ \\ \\ \prod_{i=0}^{M-1}v_i=D\;,  \end{array}\right. 
\eea
where the multi-degree of the first equation is  $(6,2,3,4,\dots,M-2,M-1,M)$.

The second Chern class of the blown-up fourfold \eqref{CY4SU} is again conveniently split in pieces
\bea\label{c2SUgroups}
c_2\left(\tilde{Z}^{SU(M)}_4\right)&=&c_2\left(\tilde{Z}^{Sp(\widetilde{M})}_4\right)+\Delta c_2^\prime\,,
\eea
where the first term is the $c_2$ of the fourfold corresponding to the parent $Sp(\widetilde{M})$ singularity and it is given by formula \eqref{c2SpSing}. The second term clearly vanishes for $M=2$, for the special case $M=3$ is
\bea\label{deltac2SU3}
\Delta c_2^{\prime\, SU(3)}&=&E_2\,\Big[-10\,c_1(B_3)+\cD+2\,E_1+E_2\Big]\,.
\eea
and, finally, has the following form for $M\geq4$
\bea
\Delta c_2^\prime&=&E_{2}\,\Big[-8\,c_1(B_3)+\mathcal{D}+E_1\Big]+\nonumber\\
&&-\sum_{i=2}^{\widehat{M}}\,E_{2i}\,\Big[2\,E_{2i-2}+E_{2i}+(7i+3)\,c_1(B_3)-\left(i^2+i+1\right)\,\cD\Big]\,, 
\eea

\section{The SU(3) case}\label{SU(3)}

We have stressed several times that F-theory configurations with $SU(3)$ gauge group are somewhat special and have to be treated separately. Moreover, we have not been able to find for them any non-trivial shift to the quantization rule of the $G_4$ flux. In this appendix we want to give an hint that, in fact, such a shift is unlikely arising for a Calabi-Yau fourfold with an $SU(3)$ singularity blown-up in the usual way \cite{Bershadsky:1996nh}, i.e. by introducing the two toric vectors $v_1,v_2$.

By inspecting the expression of $c_2(\tilde{Z}_4^{SU(3)})$ (see \eqref{c2SUgroups} and \eqref{deltac2SU3}), one realizes that this quantity is even if and only if the class $c_1E_1+D(E_1+E_2)+E_2^2$ is even. Therefore
\bea
c_2(\tilde{Z}_4^{SU(3)})&\sim&c_1E_1+D(E_1+E_2)+E_2^2 \qquad\quad {\rm mod}\; H^4(\tilde{Z}_4^{SU(3)},2\mathbb{Z})\,.
\eea
This is because all the other summand appear with an even numerical factor in front and they are Poincar\'e dual to effective 4-cycles of the Calabi-Yau fourfold. Moreover, as one can see from eq. \eqref{CY4SU}, the equation of the $SU(3)$ Calabi-Yau hypersurface contains the term $v_2Y^2$. This term is necessary, in the sense that we cannot deform it away without making the hypersurface singular\footnote{One can indeed check that, for instance, the codimension two locus $\{\sigma=X=Z=0\}$ would become singular.}. But now the two elements of the Stanley-Reisner ideal of the ambient variety \eqref{SRSU3} $v_0XY$ and $v_0Xv_2$ tell us that, on the blown-up fourfold
\bea
D(E_1+E_2)+E_2^2&\sim&E_1^2\qquad\quad {\rm mod}\; H^4(\tilde{Z}_4^{SU(3)},2\mathbb{Z})\,.
\eea
Finally, the element $v_1Y$ implies that 
\bea
E_1^2&\sim&c_1E_1\qquad\quad {\rm mod}\; H^4(\tilde{Z}_4^{SU(3)},2\mathbb{Z})\,.
\eea
While odd values of $c_2$ could still be found in principle on non-holomorphic 4-cycles, this argument nevertheless points towards the conjecture that in fact $c_2(\tilde{Z}_4^{SU(3)})$ is always an even class.

\providecommand{\href}[2]{#2}\begingroup\raggedright\endgroup


\begin{thebibliography}{10}

\bibitem{Collinucci:2008zs}
A.~Collinucci, ``{New F-theory lifts},''
\href{http://arxiv.org/abs/0812.0175}{{\tt arXiv:0812.0175 [hep-th]}}.

\bibitem{Collinucci:2009uh}
A.~Collinucci, ``{New F-theory lifts II: Permutation orientifolds and enhanced
  singularities},''
\href{http://arxiv.org/abs/0906.0003}{{\tt arXiv:0906.0003 [hep-th]}}.

\bibitem{Blumenhagen:2009up}
R.~Blumenhagen, T.~W. Grimm, B.~Jurke, and T.~Weigand, ``{F-theory uplifts and
  GUTs},'' \href{http://dx.doi.org/10.1088/1126-6708/2009/09/053}{{\em JHEP}
  {\bf 0909} (2009)  053},
\href{http://arxiv.org/abs/0906.0013}{{\tt arXiv:0906.0013 [hep-th]}}.

\bibitem{Blumenhagen:2009yv}
R.~Blumenhagen, T.~W. Grimm, B.~Jurke, and T.~Weigand, ``{Global F-theory
  GUTs},'' \href{http://dx.doi.org/10.1016/j.nuclphysb.2009.12.013}{{\em Nucl.
  Phys.} {\bf B829} (2010)  325--369},
\href{http://arxiv.org/abs/0908.1784}{{\tt arXiv:0908.1784 [hep-th]}}.

\bibitem{Grimm:2009yu}
T.~W. Grimm, S.~Krause, and T.~Weigand, ``{F-Theory GUT Vacua on Compact
  Calabi-Yau Fourfolds},''
  \href{http://dx.doi.org/10.1007/JHEP07(2010)037}{{\em JHEP} {\bf 07} (2010)
  037},
\href{http://arxiv.org/abs/0912.3524}{{\tt arXiv:0912.3524 [hep-th]}}.

\bibitem{Marsano:2009ym}
J.~Marsano, N.~Saulina, and S.~Schafer-Nameki, ``{F-theory Compactifications
  for Supersymmetric GUTs},''
  \href{http://dx.doi.org/10.1088/1126-6708/2009/08/030}{{\em JHEP} {\bf 08}
  (2009)  030},
\href{http://arxiv.org/abs/0904.3932}{{\tt arXiv:0904.3932 [hep-th]}}.

\bibitem{Chen:2010ts}
C.-M. Chen, J.~Knapp, M.~Kreuzer, and C.~Mayrhofer, ``{Global SO(10) F-theory
  GUTs},'' \href{http://arxiv.org/abs/1005.5735}{{\tt arXiv:1005.5735
  [hep-th]}}.

\bibitem{Esole:2011sm}
M.~Esole and S.-T. Yau, ``{Small resolutions of SU(5)-models in F-theory},''
\href{http://arxiv.org/abs/1107.0733}{{\tt arXiv:1107.0733 [hep-th]}}.

\bibitem{Esole:2011cn}
M.~Esole, J.~Fullwood, and S.-T. Yau, ``{$D_5$ elliptic fibrations: non-Kodaira
  fibers and new orientifold limits of F-theory},''
\href{http://arxiv.org/abs/1110.6177}{{\tt arXiv:1110.6177 [hep-th]}}.


\bibitem{Marsano:2011nn}
J.~Marsano, N.~Saulina, and S.~Schafer-Nameki, ``{On G-flux, M5 instantons, and
  U(1)s in F-theory},''
\href{http://arxiv.org/abs/1107.1718}{{\tt arXiv:1107.1718 [hep-th]}}.

\bibitem{Marsano:2010ix}
J.~Marsano, N.~Saulina, and S.~Schafer-Nameki, ``{A Note on G-Fluxes for
  F-theory Model Building},''
  \href{http://dx.doi.org/10.1007/JHEP11(2010)088}{{\em JHEP} {\bf 1011} (2010)
   088},
\href{http://arxiv.org/abs/1006.0483}{{\tt arXiv:1006.0483 [hep-th]}}.

\bibitem{Grimm:2009ef}
T.~W. Grimm, T.-W. Ha, A.~Klemm, and D.~Klevers, ``{Computing Brane and Flux
  Superpotentials in F-theory Compactifications},''
  \href{http://dx.doi.org/10.1007/JHEP04(2010)015}{{\em JHEP} {\bf 1004} (2010)
   015},
\href{http://arxiv.org/abs/0909.2025}{{\tt arXiv:0909.2025 [hep-th]}}.

\bibitem{Grimm:2011tb}
T.~W. Grimm, M.~Kerstan, E.~Palti, and T.~Weigand, ``{Massive Abelian Gauge
  Symmetries and Fluxes in F-theory},''
  \href{http://dx.doi.org/10.1007/JHEP12(2011)004}{{\em JHEP} {\bf 1112} (2011)
   004}, \href{http://arxiv.org/abs/1107.3842}{{\tt arXiv:1107.3842 [hep-th]}}.
49 pages.

\bibitem{Braun:2011zm}
A.~P. Braun, A.~Collinucci, and R.~Valandro, ``{G-flux in F-theory and
  algebraic cycles},''
  \href{http://dx.doi.org/10.1016/j.nuclphysb.2011.10.034}{{\em Nucl.Phys.}
  {\bf B856} (2012)  129--179}, \href{http://arxiv.org/abs/1107.5337}{{\tt
  arXiv:1107.5337 [hep-th]}}.
55 pages, 1 figure/ added refs, corrected typos.


\bibitem{Vafa:1996xn}
C.~Vafa, ``{Evidence for F theory},''
  \href{http://dx.doi.org/10.1016/0550-3213(96)00172-1}{{\em Nucl.Phys.} {\bf
  B469} (1996)  403--418}, \href{http://arxiv.org/abs/hep-th/9602022}{{\tt
  arXiv:hep-th/9602022 [hep-th]}}.

\bibitem{Denef:2008wq}
F.~Denef, ``{Les Houches Lectures on Constructing String Vacua},''
\href{http://arxiv.org/abs/0803.1194}{{\tt arXiv:0803.1194 [hep-th]}}.

\bibitem{Weigand:2010wm}
T.~Weigand, ``{Lectures on F-theory compactifications and model building},''
  \href{http://dx.doi.org/10.1088/0264-9381/27/21/214004}{{\em Class. Quant.
  Grav.} {\bf 27} (2010)  214004},
\href{http://arxiv.org/abs/1009.3497}{{\tt arXiv:1009.3497 [hep-th]}}.

\bibitem{Witten:1996md}
E.~Witten, ``{On flux quantization in M-theory and the effective action},''
  \href{http://dx.doi.org/10.1016/S0393-0440(96)00042-3}{{\em J. Geom. Phys.}
  {\bf 22} (1997)  1--13},
\href{http://arxiv.org/abs/hep-th/9609122}{{\tt arXiv:hep-th/9609122}}.

\bibitem{Minasian:1997mm}
R.~Minasian and G.~W. Moore, ``{K-theory and Ramond-Ramond charge},'' {\em
  JHEP} {\bf 11} (1997)  002,
\href{http://arxiv.org/abs/hep-th/9710230}{{\tt arXiv:hep-th/9710230}}.

\bibitem{Freed:1999vc}
D.~S. Freed and E.~Witten, ``{Anomalies in string theory with D-branes},'' {\em
  Asian J. Math.} {\bf 3} (1999)  819--851,
\href{http://arxiv.org/abs/hep-th/9907189}{{\tt arXiv:hep-th/9907189}}.

\bibitem{Collinucci:2010gz}
A.~Collinucci and R.~Savelli, ``{On Flux Quantization in F-Theory},''
  \href{http://dx.doi.org/10.1007/JHEP02(2012)015}{{\em JHEP} {\bf 1202} (2012)
   015}, \href{http://arxiv.org/abs/1011.6388}{{\tt arXiv:1011.6388 [hep-th]}}.
46 pages.

\bibitem{Krause:2012yh}
S.~Krause, C.~Mayrhofer, and T.~Weigand, ``{Gauge Fluxes in F-theory and Type
  IIB Orientifolds},''
\href{http://arxiv.org/abs/1202.3138}{{\tt arXiv:1202.3138 [hep-th]}}.


\bibitem{Grimm:2010ez}
T.~W. Grimm and T.~Weigand, ``{On Abelian Gauge Symmetries and Proton Decay in
  Global F-theory GUTs},'' \href{http://arxiv.org/abs/1006.0226}{{\tt
  arXiv:1006.0226 [hep-th]}}.


\bibitem{Collinucci:2008pf}
A.~Collinucci, F.~Denef and M.~Esole,
``D-Brane Deconstructions in IIB Orientifolds,''
JHEP {\bf 0902} (2009) 005
\href{http://arxiv.org/abs/0805.1573}{{\tt
  arXiv:0805.1573 [hep-th]}}.



\bibitem{Bershadsky:1996nh}
M.~Bershadsky, K.~A. Intriligator, S.~Kachru, D.~R. Morrison, V.~Sadov, {\em et
  al.}, ``{Geometric singularities and enhanced gauge symmetries},''
  \href{http://dx.doi.org/10.1016/S0550-3213(96)90131-5}{{\em Nucl.Phys.} {\bf
  B481} (1996)  215--252}, \href{http://arxiv.org/abs/hep-th/9605200}{{\tt
  arXiv:hep-th/9605200 [hep-th]}}.

\bibitem{Fulton}
W.~Fulton, ``{Intersection Theory},'' {\em Berlin, New York: Springer-Verlag}
  .

\bibitem{Andreas:1999ng}
B.~Andreas and G.~Curio, ``{On discrete twist and four flux in N=1 heterotic /
  F theory compactifications},'' {\em Adv.Theor.Math.Phys.} {\bf 3} (1999)
  1325--1413, \href{http://arxiv.org/abs/hep-th/9908193}{{\tt
  arXiv:hep-th/9908193}}.

\bibitem{Grimm:2010ks}
T.~W. Grimm, ``{The N=1 effective action of F-theory compactifications},''
  \href{http://arxiv.org/abs/1008.4133}{{\tt arXiv:1008.4133 [hep-th]}}.

\bibitem{Grimm:2011sk}
T.~W. Grimm and R.~Savelli, ``{Gravitational Instantons and Fluxes from
  M/F-theory on Calabi-Yau fourfolds},''
  \href{http://dx.doi.org/10.1103/PhysRevD.85.026003}{{\em Phys.Rev.} {\bf D85}
  (2012)  026003}, \href{http://arxiv.org/abs/1109.3191}{{\tt arXiv:1109.3191
  [hep-th]}}.
47 pages, 2 figures.

\bibitem{Donagi:2009ra}
R.~Donagi and M.~Wijnholt, ``{Higgs Bundles and UV Completion in F-Theory},''
  \href{http://arxiv.org/abs/0904.1218}{{\tt arXiv:0904.1218 [hep-th]}}.

\bibitem{Cvetic:2010rq}
M.~Cvetic, I.~Garcia-Etxebarria, and J.~Halverson, ``{Global F-theory Models:
  Instantons and Gauge Dynamics},'' \href{http://arxiv.org/abs/1003.5337}{{\tt
  arXiv:1003.5337 [hep-th]}}.

\end{thebibliography}
\end{document}